\documentclass[twocolumn]{aastex63}

\usepackage{color}
\usepackage{graphicx}
\usepackage{amsmath}
\usepackage{amssymb}
\usepackage{yfonts}
\hypersetup{pdfauthor={Name}}

\def\ra#1#2#3{#1$^{\rm h}$#2$^{\rm m}$#3$^{\rm s}$}
\def\dec#1#2#3{#1$^\circ$#2$'$#3$''$}
\def\nod{\nodata}
\def\grb{GRB\,180418A}
\def\swift{{\it Swift}}
\def\fermi{{\it Fermi}}

\accepted{}

\shorttitle{The possibly-short GRB\,180418A}
\shortauthors{Rouco Escorial et al.}

\begin{document}
\sloppy

\title{GRB\,180418A: A possibly-short GRB with a wide-angle outflow in a faint host galaxy}

\correspondingauthor{Alicia Rouco Escorial}
\email{alicia.rouco.escorial@northwestern.edu}

\newcommand{\NU}{\affiliation{Center for Interdisciplinary Exploration and Research in Astrophysics (CIERA) and Department of Physics and Astronomy, Northwestern University, 1800 Sherman Ave, Evanston, IL 60201, USA}}
\newcommand{\GSFC}{\affiliation{NASA Goddard Space Flight Center, University of Maryland, Baltimore County, Greenbelt, MD 20771, USA}}
\newcommand{\CfA}{\affiliation{Center for Astrophysics\:$|$\:Harvard \& Smithsonian, 60 Garden St. Cambridge, MA 02138, USA}}
\newcommand{\Einstein}{\altaffiliation{NASA Einstein Fellow}}
\newcommand{\NASA}{\altaffiliation{NASA Postdoctoral Fellow}}
\newcommand{\UAH}{\affiliation{Center for Space Plasma and Aeronomic Research, University of Alabama in Huntsville, 320 Sparkman Drive, Huntsville, AL 35899, USA}}
\newcommand{\USRA}{\affiliation{Science and Technology Institute, Universities Space Research Association, Huntsville, AL 35805, USA}}
\newcommand{\Arizona}{\affiliation{University of Arizona, Steward Observatory, 933 N. Cherry Avenue, Tucson, AZ 85721, USA}}
\newcommand{\Bath}{\affiliation{Department of Physics, University of Bath, Claverton Down, Bath, BA2 7AY, UK}}
\newcommand{\OU}{\affiliation{Astrophysical Institute, Department of Physics and Astronomy, 251B Clippinger Lab, Ohio University, Athens, OH 45701, USA}}
\newcommand{\Adler}{\affiliation{The Adler Planetarium, Chicago, IL 60605, USA}}
\newcommand{\GeminiN}{\affiliation{Gemini Observatory/NSF's NOIRLab, 670 N. A'ohoku Place, Hilo, HI, 96720, USA}}
\newcommand{\UMD}{\affiliation{Joint Space-Science Institute, University of Maryland, College Park, MD 20742, USA}}
\newcommand{\GWU}{\affiliation{Department of Physics, The George Washington University, Washington, DC 20052, USA}}
\newcommand{\Leicester}{\affiliation{School of Physics and Astronomy, University of Leicester, University Road, Leicester, LE1 7RH, UK}}
\newcommand{\Marin}{\affiliation{College of Marin, 120 Kent Avenue, Kentfield 94904 CA, USA}}
\newcommand{\UVI}{\affiliation{University of the Virgin Islands, \#2 Brewers bay road, Charlotte Amalie, 00802 USVI, USA}}
\newcommand{\Radboud}{\affiliation{Department of Astrophysics/IMAPP, Radboud University, 6525 AJ Nijmegen, The Netherlands}}
\newcommand{\Warwick}{\affiliation{Department of Physics, University of Warwick, Coventry, CV4 7AL, UK}}
\newcommand{\Birmingham}{\affiliation{Birmingham Institute for Gravitational Wave Astronomy and School of Physics and Astronomy, University of Birmingham, Birmingham B15 2TT, UK}}
\newcommand{\Edinburgh}{\affiliation{Institute for Astronomy, University of Edinburgh, Royal Observatory, Blackford Hill, EH9 3HJ, UK}}
\newcommand{\Caltech}{\affiliation{Cahill Center for Astrophysics, California Institute of Technology, 1200 E. California Blvd. Pasadena, CA 91125, USA}}
\newcommand{\LJMU}{\affiliation{Astrophysics Research Institute, Liverpool John Moores University, 146 Brownlow Hill, Liverpool L3 5RF, UK}}
\newcommand{\CCA}{\affiliation{Center for Computational Astrophysics, Flatiron Institute, 162 W. 5th Avenue, New York, NY 10011, USA}}
\newcommand{\Columbia}{\affiliation{Department of Physics and Columbia Astrophysics Laboratory, Columbia University, New York, NY 10027, USA}}
\newcommand{\CRESST}{\affiliation{Center for Research and Exploration in Space Science and Technology (CRESST) and NASA Goddard Space Flight Center, Greenbelt, MD 20771, USA}}
\newcommand{\Maryland}{\affiliation{Department of Physics, University of Maryland, Baltimore County, 1000 Hilltop Circle, Baltimore, MD 21250, USA}} 
\author[0000-0003-3937-0618]{A.~Rouco~Escorial} 
\NU

\author[0000-0002-7374-935X]{W.~Fong}
\NU

\author[0000-0002-2149-9846]{P.~Veres}
\UAH

\author[0000-0003-1792-2338]{T.~Laskar}
\Bath

\author[0000-0002-7851-9756]{A.~Lien}
\CRESST
\Maryland

\author[0000-0001-8340-3486]{K.~Paterson}
\NU

\author[0000-0002-4443-6725]{M.~Lally}
\NU

\author[0000-0003-0526-2248]{P.~K.~Blanchard}
\NU

\author[0000-0002-2028-9329]{A.~E.~Nugent}
\NU

\author[0000-0003-3274-6336]{N.~R.~Tanvir} 
\Leicester

\author[0000-0002-1533-9037]{D.~Cornish}
\NU


\author[0000-0002-9392-9681]{E.~Berger} 
\CfA

\author{E.~Burns}
\NASA\GSFC

\author[0000-0003-1673-970X]{S.~B.~Cenko} 
\GSFC\UMD


\author[0000-0002-9118-9448]{B.~E.~Cobb}
\GWU

\author{A.~Cucchiara} 
\Marin\UVI

\author[0000-0002-0587-7042]{A.~Goldstein}
\USRA

\author[0000-0002-8297-2473]{R.~Margutti}
\NU

\author[0000-0002-4670-7509]{B.~D.~Metzger}
\CCA\Columbia

\author{P.~Milne}
\Arizona

\author[0000-0001-7821-9369]{A.~Levan} 
\Radboud\Warwick

\author[0000-0002-2555-3192]{M.~Nicholl} 
\Birmingham\Edinburgh


\author[0000-0001-5510-2424]{Nathan~Smith} 
\Arizona

\begin{abstract}
We present X-ray and multi-band optical observations of the afterglow and host galaxy of GRB\,180418A, discovered by {\it Swift}/BAT and {\it Fermi}/GBM. We present a reanalysis of the GBM and BAT data deriving durations of the prompt emission of $T_{90}\approx$2.56\,s and $\approx 1.90$\,s, respectively. Modeling the {\it Fermi}/GBM catalog of 1405~bursts (2008-2014) in the Hardness--$T_{90}$ plane, we obtain a probability of $\approx 60\%$ that GRB\,180418A is a short-hard burst. From a combination of {\it Swift}/XRT and {\it Chandra} observations, the X-ray afterglow is detected to $\approx 38.5$~days after the burst, and exhibits a single power-law decline with $F_{\rm X} \propto t^{-0.98}$. Late-time Gemini observations reveal a faint r\,$\approx$\,25.69~mag host galaxy at an angular offset of $\approx 0.16''$. At the likely redshift range of $z \approx$\,1-2.25, we find that the X-ray afterglow luminosity of GRB\,180418A is intermediate between short and long GRBs at all epochs during which there is contemporaneous data, and that GRB\,180418A lies closer to the $E_{\gamma,{\rm peak}}-E_{\gamma,{\rm iso}}$ correlation for short GRBs. Modeling the multi-wavelength afterglow with the standard synchrotron model, we derive the burst explosion properties and find a jet opening angle of $\theta_{\rm j} \gtrsim 9-14^{\circ}$. If GRB\,180418A is a short GRB that originated from a neutron star merger, it has one of the brightest and longest-lived afterglows along with an extremely faint host galaxy. If instead the event is a long GRB that originated from a massive star collapse, it has among the lowest luminosity afterglows, and lies in a peculiar space in terms of the Hardness--$T_{90}$ and $E_{\gamma,{\rm peak}}-E_{\gamma,{\rm iso}}$ planes.

\end{abstract}


\section{Introduction}\label{sec:GRB180418A_intro}

Gamma-ray bursts (GRBs) can be divided into two classes depending on their gamma-ray duration ($T_{\rm 90}$) and hardness of their $\gamma$-ray spectra: short-hard ($T_{\rm 90} \leq 2$\,s) and long-soft ($T_{\rm 90} > 2$\,s) bursts \citep{Mazets1981,Norris1984,Dezalay1992,Kouveliotou1993}. Multi-wavelength observations of their synchrotron emission, or `afterglows' \citep[e.g.,][]{Rees1992,Meszaros1993,Paradijs2000} reveal specific information about the energetics, environments and progenitor channels of these events, as well as the features of the highly relativistic jets that are expected to be launched by the central engine \citep{Rhoads1997,Panaitescu2002,Piran2005}. Since the launches of the Neil Gehrels \textit{Swift} Observatory \citep[\textit{Swift};][]{Gehrels2004} and \textit{Fermi} Gamma-ray  Space  Telescope \citep[\textit{Fermi};][]{GLAST1999}, more than 360 GRBs with known redshifts have been detected \citep{Lien2016,vonKienlin+20GBM10yrcat}. The joint power of both observatories has yielded not only an increase in the number of detected GRBs, but also improved localizations of the events, allowing for secure associations to host galaxies.

Although the classification in terms of $\gamma$-ray hardness and $T_{90}$ encompasses the large majority of GRBs, there are some events which defy clear classification under this scheme. The lack of supernova detections for some long-duration bursts \citep[e.g., GRBs\,060505 and 060614;][]{Valle2006,Fynbo2006}, the misidentification of host galaxies yielding to the incorrect classification of GRBs \citep[e.g., GRB\,060912A;][]{Levan2007}, the longer duration of some events with similar $\gamma$-ray hardness to the short GRB population \citep[e.g., GRBs\,090607 and 100816A;][]{Barthelmy2009,DAvanzo2014}, and short-duration bursts with similar hardness and energy scales to those of the long GRBs \citep[e.g., GRBs\,090426 and 201015A;][]{Antonelli2009,Markwardt2020} reveal the ambiguous nature of certain cases and the blurred lines between the GRB populations. Other metrics based on $\gamma$-ray information exist, such as adherence to the Yonetoku/Amati relation between the $\gamma$-ray peak energy and the isotropic-equivalent $\gamma$-ray energy \citep{Amati2002,Yonetoku2004}, and data-based probability schemes \citep{Bromberg2013,Jespersen2020}. In addition to the traditional GRB classification (short and long), a few studies \citep[e.g.,][]{Horvath2006,deUgarte2011} have proposed a third group of GRBs with intermediate durations, generally with $T_{90}$ between $2-10$\,s. However, the existence of such a class has been a topic of debate since the existence of this third group depends on the instruments and the reference frames used \citep[for an in depth study see][]{Kulkarni2017}.

As a class, long GRBs have been discovered up to $z\approx9.4$ \citep[e.g.,][]{Tanvir2009,Belczynski2010,Cucchiara2011,Salvaterra2015}, with median isotropic-equivalent energies of the order of $\approx 10^{51}$\,erg \citep{Frail2001,Berger2003,Gehrels2008,Laskar2014a}. The association of long GRBs with Type Ic supernovae \citep[e.g.,][]{Galama1998,Woosley2006,Hjorth2012a}, their small offsets from their host galaxies \citep{Bloom2002,Fruchter2006,Blanchard2016}, their high circumburst densities of $\approx 0.1$--$100$\,cm$^{-3}$ \citep{Panaitescu2002,Yost2003,Laskar2018a}, and their exclusive origins from star-forming galaxies \citep{Wainwright2007}, demonstrate that long GRBs result from the deaths of massive stars. On the other hand, short GRBs are detected at much lower redshifts, $z\approx 0.1-2.2$ \citep[e.g.,][]{Fong2013,Berger2014b} as a result of a combination of observational bias and the delay time distribution from their compact object binary progenitors \citep{Selsing2018,Paterson2020}. These events are less energetic, with observed median isotropic-equivalent energies of $\approx 10^{49}$\,erg, and occur in environments with lower densities, i.e. $\approx 10^{-3}$--$10^{-2}$\,cm$^{-3}$ \citep{Nakar2007,Nicuesa2012,Berger2014a,Fong2015}, commensurate with their larger offsets from their host galaxies \citep[][]{FongBerger2013}. The discovery of the first binary neutron star (BNS) merger gravitational-wave event, GW170817 \citep{Abbott2017a} in conjunction with a short gamma-ray burst GRB\,170817A \citep{Goldstein2017,Savchenko2017}, provided direct evidence that at least some short GRBs originate from BNS mergers.

One of the most important parameters that can be gleaned from GRB afterglows is the jet opening angle, because their inference has direct consequences on the calculation of the true energy scale and rates of these events \citep[e.g.,][]{Frail2001,Fong2015,Mandhai2018}. For on-axis orientations, the jet opening angles can be determined from the detection of sudden steepenings in the broad-band afterglow light curves \citep[also called `jet breaks';][]{Piran1999,Rhoads1999,Sari1999,Panaitescu2005}, while limits on the jet opening angles can be inferred from the lack of jet breaks in the light curves. X-ray observations have played a leading role in these studies, thanks to the rapid X-ray detections and follow-up of most GRBs provided by {\it Swift} \citep{Evans2007,Evans2009,Nysewander2009,Racusin2009}. This facilitates not only the determination of the GRB afterglow decay rates, but also a tightening of the constraints on the limits of the jet opening angles.

The relative brightness of long GRB afterglows \citep[e.g.,][]{Bernardini2012,delVecchio2016} has led to the successful identification of jet breaks in their light curves, with opening angles of $<10^{\circ}$ \citep{Frail2001,Racusin2009,Kann2010,Ryan2015,Goldstein2016}. However, short GRB afterglows are generally fainter \citep[e.g.,][]{Gehrels2008,Nysewander2009,Kann2010,Berger2013,Fong2015}, making the identification of jet breaks in their light curves more challenging. For only a few short GRBs, jet opening angles have been measured between $\approx 2-7^{\circ}$ \citep[e.g.][]{Burrows2006,Soderberg2006,Fong2012,Troja2016,Lamb2019} while for the remaining events, meaningful lower limits of $\gtrsim 4-25^{\circ}$ have been inferred at $\gtrsim2$~days after the trigger \citep{Fong2015,Jin2018}. This may indicate that short GRBs have wider jets than their long-duration counterparts. So far, there is no clear mechanism to keep the jet collimated in the case of short GRBs \citep{Sari1999,MeszarosRees2001,Zhang2003}.

In this paper, we present the multi-wavelength afterglow monitoring of the potentially short GRB 180418A, spanning the X-ray and optical bands, and the discovery of its host galaxy. Our late-time \textit{Chandra} detections of the X-ray afterglow of GRB\,180418A extend up to $\approx 38.5$~days after the trigger of the burst, representing one of the latest X-ray detections of a potential short GRB. In Section\,\ref{sec:GRB180418A_discovery}, we present the burst discovery and the \textit{Swift} and \textit{Fermi} data re-analysis classification of the burst. In Section\,\ref{sec:GRB180418A_Followup}, we introduce the multi-wavelength afterglow observations and discovery of the host galaxy. We discuss the burst explosion properties and limits on the jet opening angle in Section\,\ref{sec:GRB180418A_analysis}. We compare our X-ray results to the \textit{Swift} GRB population with known redshifts in Section\,\ref{sec:GRB180418A_ComparisonPopulation}. In Section\,\ref{sec:GRB180418A_discussion}, we discuss our results in the context of the short and long GRB populations. Finally in Section\,\ref{sec:GRB180418A_conclusions}, we summarize our conclusions.

Unless mentioned otherwise, all observations are reported in AB mag and have been corrected for Galactic extinction  in the  direction  of  the  burst \citep{Schlafly2011}. The cosmology employed in this paper is standard, with H$_0=69.6$\,km~s$^{-1}$~Mpc$^{-1}$, $\Omega_{\rm{M}}=0.286$, $\Omega_{\rm{vac}}=0.714$ \citep{Bennett2014}.

\section{Burst Discovery and Classification}
\label{sec:GRB180418A_discovery}

\subsection{Initial Observations of \grb}
\label{sec:GRB180418A_detection}

\grb\ triggered the Burst Alert Telescope (BAT; \citealt{Barthelmy2005}) on-board \swift\ at 06:44:06 UT on 2018 April 18, with an initially reported duration of $\sim$1.5\,s \citep{Delia2018}. \swift/BAT located the GRB to a refined position of RA\,(J2000)$=$\ra{11}{20}{31.6} and Dec\,(J2000)$=+$\dec{24}{55}{28.9} ($1.2'$ radius uncertainty, 90\% confidence; \citealt{Delia2018}) and revealed a single peaked light curve with $T_{90}=2.29\pm0.83$\,s in the $15-350$\,keV energy band \citep{Palmer2018}. Additionally, \grb\ independently triggered and was detected by the Gamma-ray Burst Monitor (GBM; \citealt{Meegan2009}) aboard \fermi\ at 06:44:06.28 UT. The GBM light curve consists of a single peak with an initially reported duration of $T_{90}\sim 2.5$\,s in the $50-300$\,keV energy range \citep{Bissaldi2018}.

The \swift/X-ray Telescope (XRT; \citealt{Burrows2005}) started the follow-up of \grb\ at $\delta \rm t \approx 3.88\times10^3$\,s (where $\delta$t represents the elapsed time since the BAT trigger). The slight delay of XRT observations with respect to the BAT detection was due to an observing constraint (\citealt{Delia2018}). An uncatalogued X-ray source was discovered within the BAT position with an enhanced XRT position of RA\,(J2000)$=$\ra{11}{20}{29.17} and Dec\,(J2000)$=+$\dec{24}{55}{59.1} ($1.8''$ radius uncertainty, 90\% confidence; \citealt{Goad2018}), and identified as the X-ray afterglow of \grb. The afterglow of GRB\,180418A was also detected and monitored in the optical band by ground-based facilities (see Section\,\ref{sec:GRB180418A_opt_obs}) and the \swift\ Ultraviolet/Optical Telescope (UVOT; \citealt{Roming2005}). UVOT detected the optical afterglow of \grb\ in the $white$, $u$, $uvw1$, and $uvm2$ filters ($\sim 19.0-19.5$\,mag at $\delta \rm t \approx 3.5 \times10^3$\,s), but yielded non-detections in the $v$, $b$ and $uvw2$ filters \citep{Siegel2018}. 

In addition, radio observations of the field of \grb\ were performed using the Arcminute Microkelvin Imager (AMI; \citealt{Zwart2008}) Large Array at 15.5 GHz at $\delta \rm t \approx 0.61$, 2.61 and 4.58 days. The radio afterglow was not detected to $3\sigma$ upper limits of $\gtrsim99\,\mu$Jy, $\gtrsim81\,\mu$Jy and $\gtrsim93\,\mu$Jy, respectively \citep{Bright2018}.

%
\begin{figure}
	\centering
	\includegraphics[width=\columnwidth]{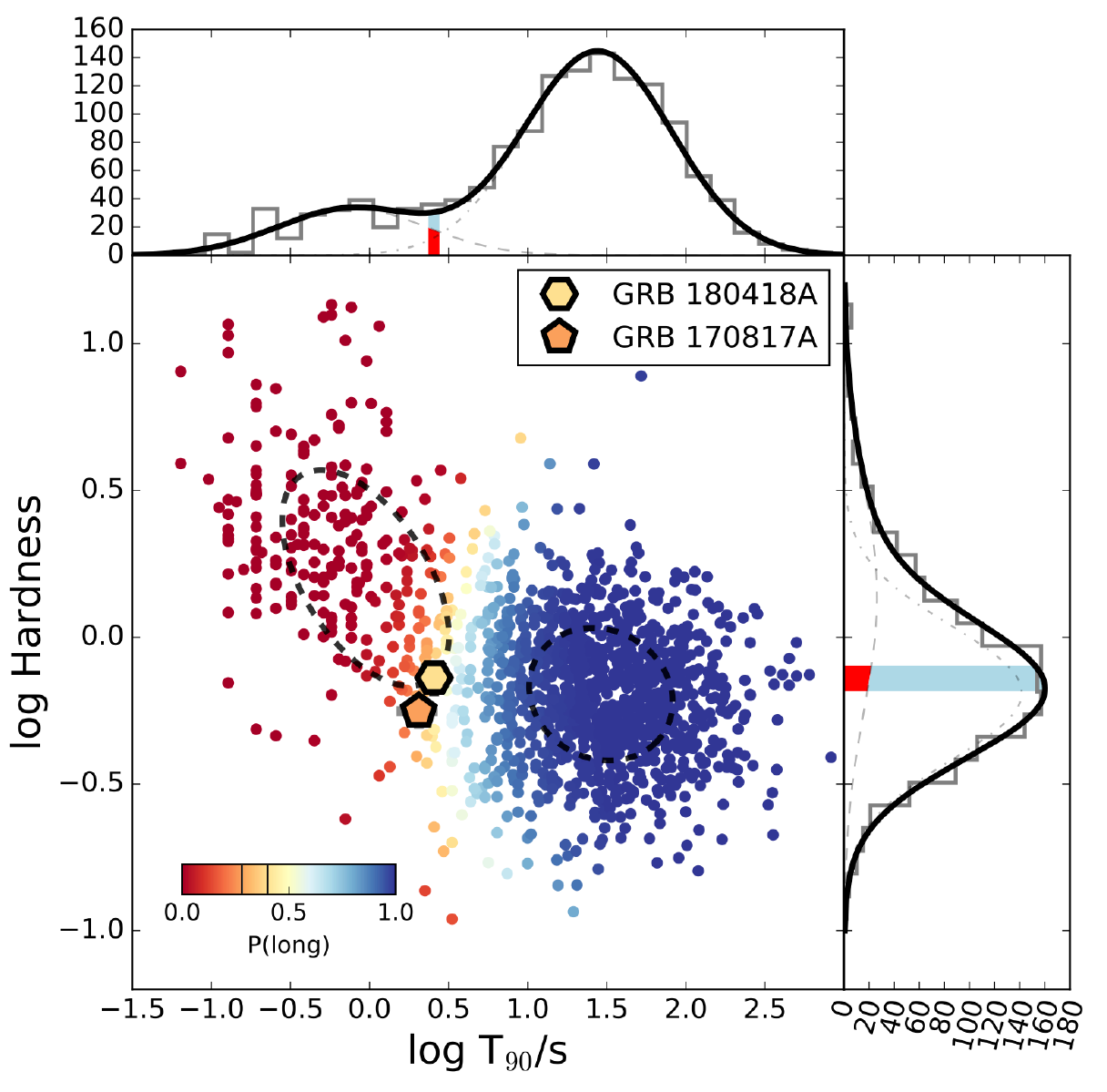}
	\vspace{-0.23in}
    \caption{The Hardness-$T_{90}$ (observer frame) plane of 1405 bursts detected by \textit{Fermi}/GBM \citep{Bhat2016}. The color scale from red to blue indicates the probability that a given event is a long GRB (P$_{\rm long}$), where a value of P$_{\rm long}=0$ indicates a short GRB. In the projected histograms for each parameter (top and right), the contributions from the short (red) and long (light blue) GRB populations at the position of GRB\,180418A are shown, containing the classification based only on either duration or hardness alone. We use two 2-dimensional Gaussians to fit the distributions, where dashed-line ellipses correspond to $1\sigma$ confidence. The hexagon and pentagon indicate the locations of GRB\,180418A and GRB\,170817A, respectively. The position of each source has also been indicated with black lines in the probability color bar.}
    \label{fig:180418a_Fermi}
\end{figure}
%

\subsection{Classification of \grb}
\label{sec:GRB180418A_classification}

The initial reported duration of \grb\ was $T_{\rm 90}$ $\sim1.5$--$2.5$\,s \citep{Delia2018,Bissaldi2018}. This makes the immediate classification of \grb\ ambiguous, given that the traditional division between short and long GRBs is placed at $T_{90} \sim 2$\,s (e.g., \citealt{Kouveliotou1993}), and the exact location of this division is detector-dependent \citep{Bromberg2013,Lien2016,vonKienlin+20GBM10yrcat}. Thus, to clarify the classification on this burst, we reanalyze the available \fermi/GBM and \swift/BAT data to determine both the duration and hardness ratio of \grb.

First, we analyze the \swift/BAT data of \grb, building upon previous analyses reported by \citet{Palmer2018} and \citet{Becerra2019}. The \swift/BAT catalog\footnote{\url{https://swift.gsfc.nasa.gov/results/batgrbcat/GRB180418A/web/GRB180418A.html}} reports a value of $T_{90}=4.41\pm2.49$\,s ($15-350$\,keV), which is calculated using the standard set-up of the BAT pipeline with a bin size of $4$\,ms (\texttt{batgrbproduct}; \citealt{Lien2016}). We re-analyzed the BAT light curve using two additional standard bin sizes of $16$\,ms and $64$\,ms, which give $T_{90}=1.90\pm0.76$\,s, and $1.92\pm0.72$\,s, respectively. Although the $T_{90}$ values of all three bin choices are consistent with each other within the errors, we find that the duration and larger uncertainty obtained by the 4\,ms-binned light curve may be reflective of a potential weak tail emission following the initial peak of the light curve. In order to investigate this possibility, we create an image for $\delta \rm t=2.0-4.0$\,s ($15-350$\,keV) and detect the burst at $\sim 2.9\sigma$ level; therefore, we cannot rule out that the emission during this interval is due to a noise fluctuation.

Adopting a value of $T_{90}=1.90\pm0.76$\,s ($16$\,ms bin), we calculate the \grb\ $\gamma$-ray fluence, $f_\gamma$, and hardness ratio following the same procedure used in the third \swift/BAT catalog \citep{Lien2016}. The spectrum corresponding to this $T_{90}$ value is best-fit by a single power-law model, $f(E) \propto E^{\Gamma_{\gamma, {\rm PL}}}$, (following the criteria in \citealt{Sakamoto2011}) with a photon index ($\Gamma_{\gamma, {\rm PL}}$) of $\approx-1.45$. We measure a fluence of $f_{\gamma} = (2.85\pm0.20)\times10^{-7}$\,erg~cm$^{-2}$ ($15-350$\,keV) and hardness ratio, defined as $f_{\gamma}$($50-100$\,keV) / $f_{\gamma}$($25-50$\,keV), of 1.47. In the context of the Hardness-$T_{90}$ plane for \swift/BAT GRBs \citep{Lien2016}, \grb\ appears to be a limiting case and close to the dividing threshold between short and long GRBs (although more recent machine learning schemes based on the {\it Swift}/BAT catalog data alone classify GRB\,180418A as ``long''; \citealt{Jespersen2020}).

Next, we analyze the \fermi/GBM data, in which the $T_{90}$ duration is typically measured in the $50-300$\,keV energy range. The total flux, and thus $T_{90}$ value, is obtained by using the \texttt{RMFIT} software to fit the background-subtracted spectrum for each time bin with an exponential cutoff power-law model \citep{Gruber2014,Bhat2016}, and a default temporal bin resolution of $64$\,ms (post-trigger resolution of the \texttt{CTIME} data type). Employing this method for \grb, we measure a single-peaked light curve with a duration of $T_{90}=2.56\pm0.20$\,s ($1\sigma$ errors) in the $50-300$\,keV energy range, and calculate the burst hardness over $T_{90}$ as the ratio of deconvolved counts in the $50-300$\,keV to $10-50$\,keV energy ranges \citep{vonKienlin+20GBM10yrcat}. The resulting GBM hardness ratio of \grb\ is $0.728\pm0.074$. We also find that the best-fit model of the burst spectrum is a comptonized model (\texttt{COMP}; an exponentially cutoff power law) characterized by $\Gamma_{\rm \gamma,COMP}=-1.20\pm0.15$ and a peak energy ($E_{\rm peak}$) of $329\pm123$\,keV (C-stat$=24.08$ and d.o.f$=16$ using Castor statistics; \citealt{Dorman2003,Ackermann2011}). We derive $f_{\gamma}=9.03 \times 10^{-7}$\,erg~cm$^{-2}$ ($10-1000$\,keV) over the interval duration. Adopting the parameters of the \texttt{COMP} model and fiducial redshift ($z$) values of 1.0 and 1.5, we obtain the $1-10000$\,keV isotropic-equivalent $\gamma$-ray energy ($E_{\gamma, \rm iso}$) values of $2.71\times10^{51}$\,erg and $5.95 \times 10^{51}$\,erg, respectively.

We compare the values for \grb\ to those of the \textit{Fermi}/GBM catalog comprising 1405 GRBs (2008 July 12 to 2014 July 11), which were analyzed in the same manner (\citealt{Bhat2016}; see Figure~\ref{fig:180418a_Fermi}). We use the position of \grb\ in the Hardness-$T_{90}$ plane to quantify the probability that \grb\ is a long GRB (P$_{\rm long}$). We fit the Hardness-$T_{90}$ distribution with two 2-dimensional Gaussian components using the \texttt{mclust} package \citep{Scrucca2016}. The two components of each Gaussian correspond to the short and long GRB populations. Under this scheme, the probability P$_{\rm{long}}$ can be assigned to each burst based on its location in this plane (Figure~\ref{fig:180418a_Fermi}). For \grb\, we obtain a probability P$_{\rm{long}} = 0.4$ (or conversely, $P_{\rm short}=0.6$). For comparison, we note that GRB\,170817A, which was associated with GW170817, had a shorter duration but was slightly softer, with a value of P$_{\rm{long}} = 0.28$ (\citealt{Goldstein2017}; Figure~\ref{fig:180418a_Fermi}). Based on the the value of $P_{\rm long}$ for GRB\,180418A and its similar position to GRB\,170817A in the Hardness-$T_{90}$ plane, it is more plausible that \grb\ is likely a short-hard GRB.

%
\begin{figure*}[t]
	\centering
	\includegraphics[scale=0.4]{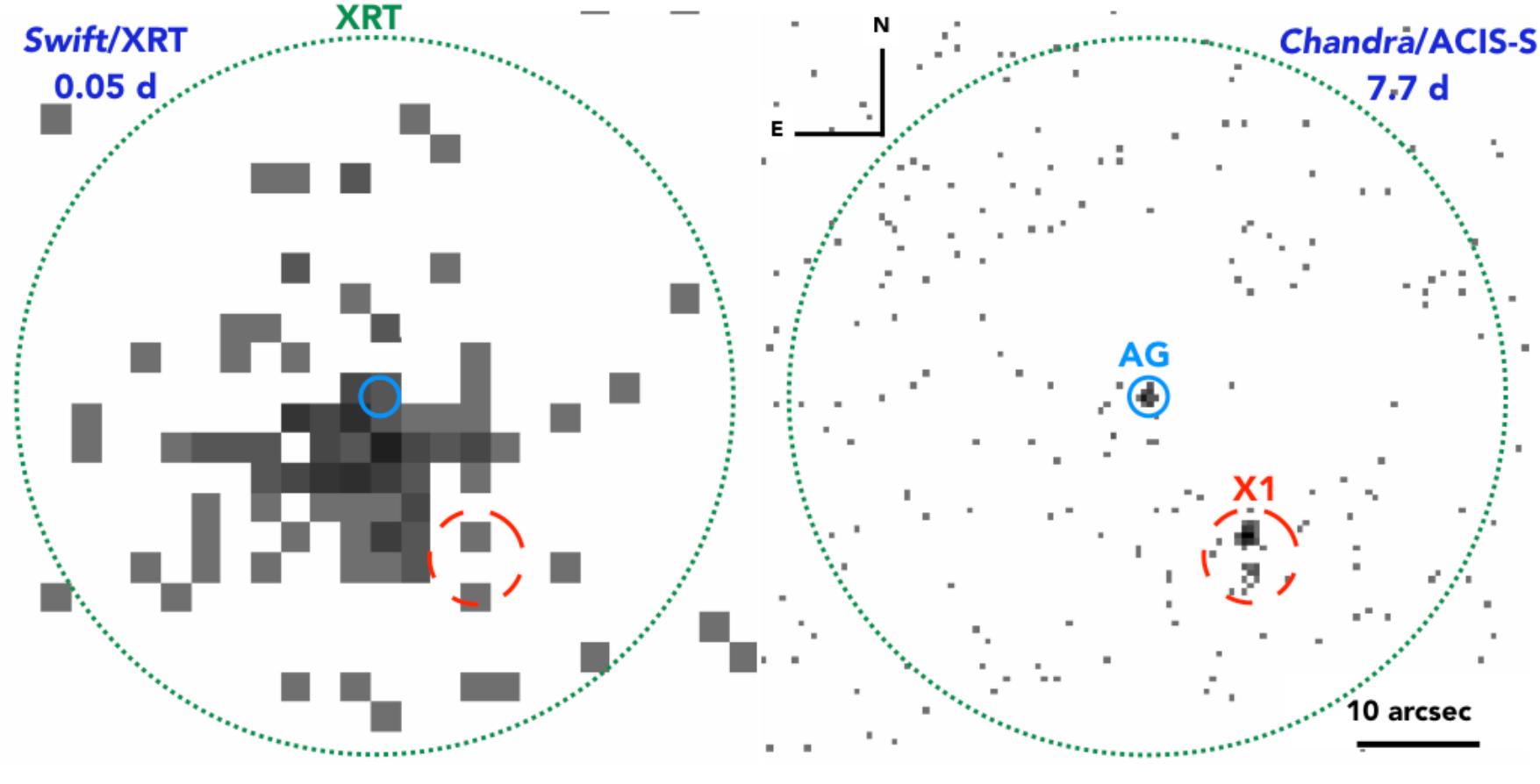}
    \caption{X-ray imaging of the \textit{Swift}/XRT (left) and \textit{Chandra}/ACIS-S (right) images of GRB\,180418A in the $0.5-8$\,keV energy bands. The large dotted green circle indicates the XRT source subtraction region while the blue circle shows the $3\sigma$ source region from the \textit{Chandra} observation. The small dashed red region in both images shows X1, which adds an extra contribution to the count rate extracted from the XRT source region.}
    \label{fig:180418a_Xray_image}
\end{figure*}
%

\section{Follow-up observations of GRB\,180418A}\label{sec:GRB180418A_Followup}

\subsection{X-ray Observations}\label{sec:GRB180418A_xray_observations}

\subsubsection{\textit{Chandra} Afterglow Detections}\label{sec:GRB180418A_Chandra}

%
\begin{deluxetable*}{lcccc}
\tabletypesize{\small}
\tablecaption{X-ray observations of GRB\,180418A \label{tab:GRB180418A_xray_observations}}
\tablewidth{0pt}
\tablehead{
\colhead{ObsID} & \colhead{$\delta$t} & \colhead{Exposure Time} & \colhead{$\Gamma_{\rm{X}}$} & \colhead{F$_\text{X}$}\\
\colhead{} & \colhead{(s)} & \colhead{(s)} & & \colhead{(erg~cm$^{-2}$~s$^{-1}$)}
}
\startdata
\multicolumn{5}{c}{\textit{Swift}/XRT} \\
\hline
0082642800[0] & $3.88 \times 10^3$ & $1.73 \times 10^3$ & $2.02^{+0.29}_{-0.16}$ & $3.62(^{+0.33}_{-0.31}) \times 10^{-12}$ \\
+[1]  & $1.68 \times 10^4$ & $4.95 \times 10^3$ & $1.70^{+0.22}_{-0.21}$ & $8.04(^{+1.03}_{-0.96}) \times 10^{-13}$\\
+[2]  & $1.21 \times 10^5$ & $5.10 \times 10^3$ & $1.3^{+3.2}_{-1.7}$ & $8.1(^{+5.7}_{-4.7}) \times 10^{-14}$\\
+[3]  & $5.99 \times 10^4$ & $4.69 \times 10^3$ & $1.50^{+0.61}_{-0.60}$ & $2.17(^{+0.69}_{-0.60}) \times 10^{-13}$\\
+[4-5]  & $2.02 \times 10^5$ & $9.09 \times 10^3$ & $1.39^{+1.11}_{-0.85}$ & $9.0(^{+4.0}_{-3.5}) \times 10^{-14}$\\
+[6-7]  & $6.00 \times 10^5$ & $1.09 \times 10^4$ & $2.5^{+6.7}_{-1.1}$ & $4.4(^{+2.5}_{-2.1}) \times 10^{-14}$\\
+[8]  & $1.84 \times 10^6$ & $4.42 \times 10^3$ & $"$ & $<1.2 \times 10^{-13}$\\
\cline{1-5}
\multicolumn{5}{c}{Afterglow, \textit{Chandra}/ACIS-S} \\
\hline
20180 & $6.63 \times 10^5$ & $2.41 \times 10^4$ & $2.66^{+1.00}_{-0.73}$ & $2.40(^{+0.48}_{-0.43})\times10^{-14}$\\
20181 & $1.67 \times 10^6$ & $9.80 \times 10^3$ & $"$ & $<2.6 \times 10^{-14}$\\
21092 & $3.33 \times 10^6$ & $2.76 \times 10^4$ & $"$ & $5.3(^{+2.7}_{-2.0}) \times 10^{-15}$\\
\cline{1-5}
\multicolumn{5}{c}{X1, \textit{Chandra}/ACIS-S} \\
\hline
2018[0-1] \& 21092 & $-$ & $-$ & $1.94^{+0.23}_{-0.17}$ & $6.24(^{+0.42}_{-0.40}) \times 10^{-14}$\\
\enddata
\tablecomments{The elapsed time between the trigger of the burst and the observation is given by $\delta$t. The effective exposure times (after the data were filtered for background flares) are displayed in this table. The Galactic absorption column density (N$_\text{H, MW}$) was fixed to $9.76\times10^{19}$\,cm$^{-2}$ (\citealt{HI4PICollaboration2016}) during the spectral fitting process. Spectral photon indices ($\Gamma_{\text{X}}$) were obtained in the $0.5-8$\,keV energy range, while the unabarsobed X-ray fluxes (F$_\text{X}$) were calculated for the $0.3-10$\,keV band. Confidence intervals are 1$\sigma$. The 3$\sigma$ flux upper limits were determined following the method described in Section~\ref{sec:GRB180418A_spectra}}
\end{deluxetable*}
%

We used \textit{Chandra} to obtain observations of the afterglow of GRB\,180418A using the ACIS-S detector (\citealt{Garmire2003}) at $\delta t \approx 7.7$, $19.3$ and $38.5$ days, respectively (Figure~\ref{fig:180418a_Xray_lc}; Program 19400201, PI: Fong). To reduce and analyze the data, we used the \texttt{CIAO} software package (v.\,4.12; \citealt{Fruscione2006}) and the calibration database files (\texttt{caldb}; v.\,4.9.0). We reprocessed the data to obtain new Level II event files, and filtered each observation to exclude intervals of high background activity.

For the first \textit{Chandra} observation at $\delta t \approx 7.7$~days (effective exposure time of $\sim 24$\,ks), we performed blind source detection using the \texttt{CIAO} routine \texttt{wavdetect} and detected the X-ray afterglow of GRB~180418A at a position of RA~(J2000)$=$\ra{11}{20}{29.21} and Dec~(J2000)$=+$\dec{24}{55}{59.21}, with a total positional uncertainty of $0.81''$ (combining the afterglow centroid uncertainty of $0.091''$ and the \textit{Chandra} absolute astrometric uncertainty of $0.8''$). The \textit{Chandra} position is fully consistent with the enhanced XRT position (Figure~\ref{fig:180418a_Xray_image}). From the \texttt{wavdetect} output at the \textit{Chandra} afterglow position, we obtain a total net source counts of $31\pm6$ in $\sim 24$\,ks and derive a source significance of $5\sigma$. We analyzed the two remaining observations at $19.3$ and $38.5$~days in a similar manner; a blind search yields a non-detection and a detection of $6 \pm 3$~counts in 28~ks (2$\sigma$) at the position of the afterglow, respectively.

We also detect a neighboring X-ray source complex (hereafter X1) at an angular distance of $\sim 14.76''$ from the afterglow, at RA~(J2000)$=$\ra{11}{20}{28.61} and Dec~(J2000)$=+$\dec{24}{55}{46.7} (Figure~\ref{fig:180418a_Xray_image}). While our \textit{Chandra} observations can separate these two sources, the contributions to the X-ray flux from both the afterglow and X1 are indistinguishable in XRT observations (Figure~\ref{fig:180418a_Xray_image}). Thus, we extract information from the position of X1 to model its X-ray spectral behavior and account for it in the XRT spectral analysis and derivation of the full X-ray afterglow light curve (Section~\ref{sec:GRB180418A_spectra}).

\subsubsection{\textit{Swift}/XRT Analysis}\label{sec:GRB180418A_Swift}

Given that X1 contaminates the afterglow position, we revise the \textit{Swift}/XRT data of GRB\,180418A to account for the contribution of flux from X1. We note that the automatic analysis of GRB\,180418A\footnote{\url{https://www.swift.ac.uk/xrt_curves/00826428/}} exhibits a flattening at $\delta t > 10^5$\,s, most likely an indication of contamination from X1. The XRT observations span $\delta t=3.88\times10^3-1.84\times10^6$\,s, after which the flux fades below the XRT sensitivity limit (Figure~\ref{fig:180418a_Xray_lc}; \citealt{Evans2007,Evans2009}). In total, nine XRT observations of the burst were obtained in PC mode (see Table~\ref{tab:GRB180418A_xray_observations}).

We downloaded all of the XRT observations from the \texttt{HEASARC} archive. For the reduction of the XRT data, we used the \texttt{HEASoft} software (v.6.17; \citealt{Blackburn1999,NASA2014}) and \texttt{caldb} files (v.\,20170831). We produced new event files centered on the \textit{Chandra} afterglow position utilizing the \texttt{xrtpipeline} tool and used them to perform the spectral analysis.

\subsubsection{Joint X-ray Spectral Analysis}\label{sec:GRB180418A_spectra}

%
\begin{figure}
	\centering
	\includegraphics[width=\columnwidth]{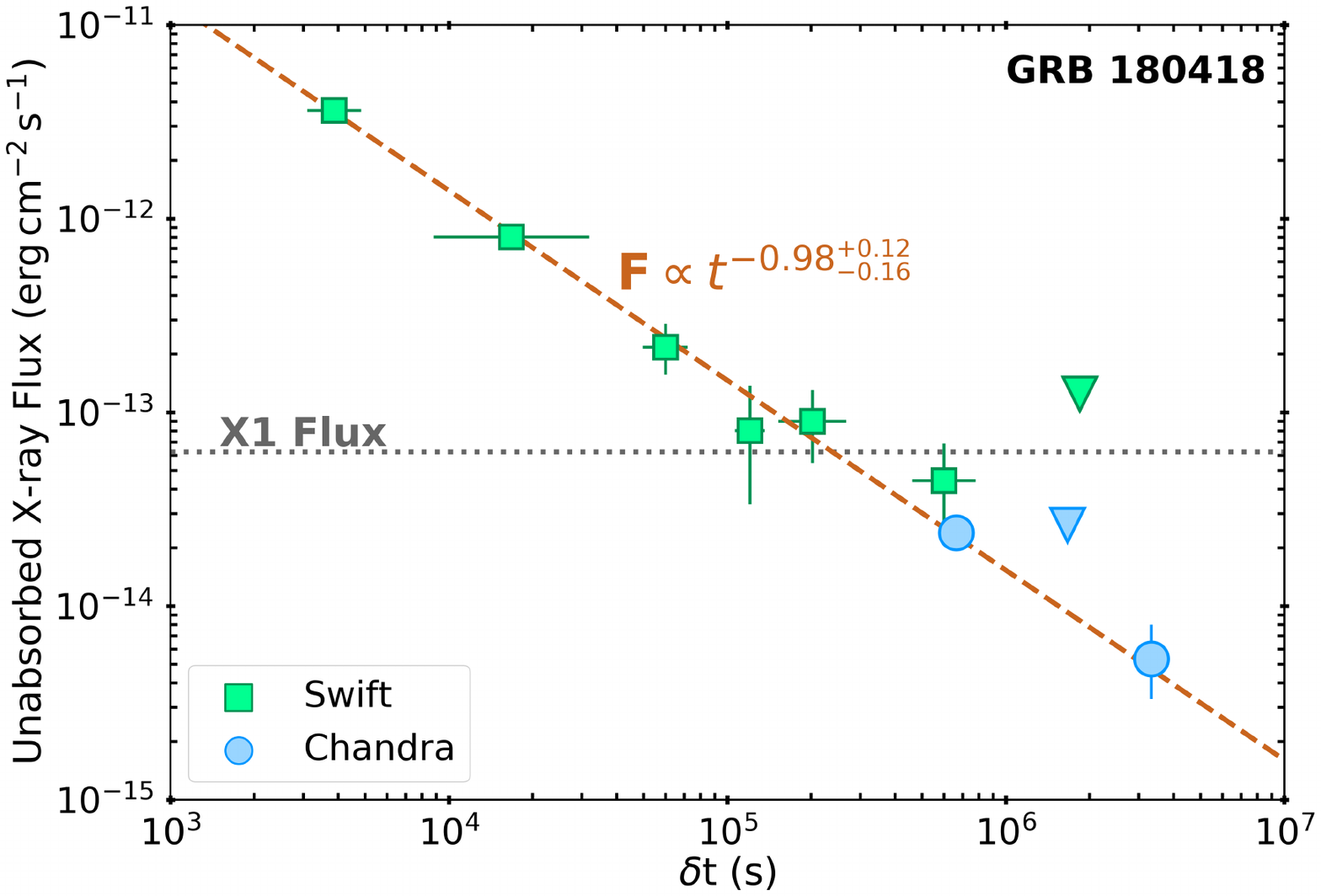}
    \caption{The \textit{Swift}/XRT (green squares) and \textit{Chandra}/ACIS-S (blue circles) unabsorbed X-ray flux light curve ($0.3-10$\,keV) for GRB\,180418A. Each observation is log-centered with the time errors denoting the duration of each observation. The flux errors are $1\sigma$. In some cases, the symbols are bigger than the errors. The \textit{Swift} and \textit{Chandra} $3\sigma$ upper limits are indicated with green and blue arrows, respectively. The dashed light brown line represents the best-fit single power-law model with $\alpha_X = -0.98^{+0.12}_{-0.16}$. The dotted horizontal grey line shows the unabsorbed X-ray flux level of X1, $\sim6.24\times10^{-14}$\,erg~cm$^{-2}$~s$^{-1}$ (see Table~\ref{tab:GRB180418A_xray_observations}).}
    \label{fig:180418a_Xray_lc}
\end{figure}
%

First, we determine the count rates of the afterglow of GRB\,180418A and X1 in the \textit{Chandra} observations. We then obtain the spectral parameters of both sources and used them to revise the XRT light curve. To obtain the {\it Chandra} count rates, we use a circular region with a radius of $1.5''$ centered on the \textit{Chandra} afterglow position, and obtain the background from a source-free annulus with inner and outer radii of $18.5''$ and $34''$, respectively. Using \texttt{CIAO/dmextract} we obtain the afterglow net count rate of $(1.26\pm0.24) \times 10^{-3}$\,counts s$^{-1}$ and $(2.0\pm1.0) \times 10^{-4}$\,counts s$^{-1}$ from the first and third \textit{Chandra} observations, respectively (see Table~\ref{tab:GRB180418A_xray_observations}). For X1, we used a circular region with a radius of $3.5''$, adjusted to encompass the entire complex (Figure~\ref{fig:180418a_Xray_image}), and a background annulus of the same size as that used for the afterglow. We generate the source and background spectra for both the afterglow and X1, as well as the necessary ancillary response file (\texttt{arf}) and redistribution matrix file (\texttt{rmf}) utilizing the \texttt{CIAO specextract} tool.

We first determine the spectral parameters of the {\it Chandra} observation at $\delta t \approx 7.7$ days, by fitting a model using \texttt{Xspec} (v.12.9.0; \citealt{Arnaud1996}) for the spectrum within the $0.5-8$\,keV energy band. We choose a bin size to ensure at least one count per bin using the \texttt{grppha} task, and to avoid any bin with negative net values when subtracting the background. In addition, we set the abundances to \texttt{WILM} (\citealt{wilms2000}), the X-ray cross-sections to \texttt{VERN} (\citealt{Verner1996}) and the statistics to W-statistics (statistics for background-subtracted Poisson data; \citealt{Wachter1979}). We employ a power-law model (\texttt{pow}) with  two absorption components (\texttt{tbabs}), i.e. \texttt{tbabs x tbabs x pow} in \texttt{Xspec}, which represent the Galactic column density (N$_{\rm{H, MW}}$) and the intrinsic absorption value (N$_{\rm{H, int}}$). We fix the Galactic contribution to N$_{{\rm{H, MW}}}=9.76\times10^{19}$\,cm$^{-2}$ (\citealt{HI4PICollaboration2016}), while leaving the rest of the spectral parameters (including N$_{\rm{H, int}}$) free. The best-fit {\it Chandra} spectrum is characterized by a spectral photon index, $\Gamma_{\rm{X}}=2.66^{+1.00}_{-0.73}$ and N$_{\rm{H, int}}<5.4\times10^{21}$\,cm$^{-2}$ ($1\sigma$ confidence). To obtain the unabsorbed flux (F$_{\rm{X}}$), we fix the spectral parameters to the best-fit values and use the convolution model, \texttt{cflux}, setting the energy range to 0.3$-$10\,keV (Table~\ref{tab:GRB180418A_xray_observations}). We repeat this procedure for the {\it Chandra} observation at $\delta t \approx 38.5$ days. The results of our spectral fits are listed in Table~\ref{tab:GRB180418A_xray_observations} and the unabsorbed fluxes are displayed in Figure~\ref{fig:180418a_Xray_lc}.

We model the spectrum of X1 in the same manner as described above in all three {\it Chandra} observations, but instead employ a single absorbed power-law model (\texttt{tbabs x pow}) in \texttt{Xspec}, as the consideration of the individual $N_{\rm H}$ contributions are not important here. We find that the spectral parameters of X1 at each epoch are consistent within $1\sigma$ errors and thus do not exhibit any significant evidence for spectral evolution between the three observations. Therefore, we link the spectral parameters between the three spectra of X1 and fit them simultaneously to better constrain the spectral shape of X1. We find the best-fit power-law spectrum of X1 is characterized by $\Gamma_{\rm X,X1} = 1.94^{+0.23}_{-0.17}$ and N$_{\rm H, X1}<1.8\times10^{21}$\,cm$^{-2}$ ($1\sigma$ confidence intervals). 

To perform the spectral analysis of all nine \textit{Swift}/XRT observations, we first combine the last four XRT observations in two groups (ObsIDs 0082642800[4-5] and 0082642800[6-7]) to ensure better statistics, resulting in seven epochs. We then obtain the spectrum of the afterglow for each observation utilizing the \texttt{Xselect} tool. For that, we use a circular source extraction region with a radius of $\sim 28.28''$ centered at the afterglow position, and a background annulus with inner and outer radii of $\sim 141.44''$ and $\sim 259.30''$, respectively, centered at the \textit{Chandra} afterglow position. We use \texttt{grppha} again for binning our spectra in order to obtain a minimum of one count per bin. For each observation, we create the exposure maps with \texttt{xrtpipeline}, the \texttt{arf} files with the \texttt{xrtmkarf} tool, and use the \texttt{rmf} files (v.14) for the spectral fitting.

We employ a two-component model to account for the combined presence of the afterglow and X1, using double and single-absorbed power-law models respectively. We use the \texttt{constant} multiplicative model to account for the cross-calibration between \textit{Swift}/XRT-PC and \textit{Chandra}/ACIS-S3. We set the XRT-PC constant value to 1 and calculate the ACIS-S3 constant value (\texttt{const}$_{\text{ACIS-S3}}=1.147$) using Table~5 from \citet{Plucinsky2017}. To specifically obtain the unabsorbed fluxes from the afterglow in the $0.3-10$\,keV energy band, we set \texttt{cflux} only for the spectral component of the model that accounts for the afterglow as follows: \texttt{(tbabs x tbabs x const x cflux x pow)$_{\text{AG}}$ + (tbabs x const x pow)$_{\text{X1}}$} in \texttt{Xspec}. The best-fit spectral parameters and unabsorbed fluxes with $1\sigma$ uncertainties are listed in Table~\ref{tab:GRB180418A_xray_observations} and shown in Figure~\ref{fig:180418a_Xray_lc}. We only find significant adjustments to the XRT fluxes relative to the automatic pipeline values for the last three epochs, as the afterglow flux approaches the level of X1.

Finally, to derive upper limits from the XRT and \textit{Chandra} observations where the afterglow is not detected, we extract the photons from the correspondening circular source regions centered on the afterglow \textit{Chandra} position using \texttt{Xselect} and \texttt{CIAO/dmextract} tools, respectively. For the \textit{Swift} observation at $\delta t \approx 21$~days, only 3 source photons are detected in $\sim 4.4$\,ks, while the same number of photons is obtained in $\sim 9.8$\,ks of \textit{Chandra} observations at $\delta t \approx 19$~days. We use Poissonian confidence levels for small numbers of X-ray events according to \citet{Gehrels1986} to calculate the $3\sigma$ count-rate upper limits and estimate the $3\sigma$ X-ray flux upper limits with WebPIMMS tool\footnote{\url{https://heasarc.gsfc.nasa.gov/cgi-bin/Tools/w3pimms/w3pimms.pl}} utilizing the best-fit spectral parameters of the first \textit{Chandra} detection. These values are listed in Table~\ref{tab:GRB180418A_xray_observations}.

%
\begin{deluxetable*}{lccccccc}[!t]
\tabletypesize{\small}
\tablecolumns{8}
\tablewidth{0pc}
\tablecaption{Afterglow and Host Galaxy Photometry of GRB\,180418A
\label{tab:gb_obs}}
\tablehead{
\colhead {Date}	 &
\colhead {$\delta t$}	 &
\colhead {Telescope} &		
\colhead {Filter} &
\colhead {Exp. Time}		    &
\colhead {Afterglow}		    &
\colhead {Host Galaxy}		    &
\colhead {A$_{\lambda}$}		   		    \\
\colhead {(UT)}	&
\colhead {(d)}	 &
\colhead {} 		&
\colhead {}	 &
\colhead {(s)}	&
\colhead {(AB~mag)}	&
\colhead {(AB~mag)}	&
\colhead {(AB~mag)}	
}
\startdata
2018 Apr 18.410 & 0.13 & Gemini-N/GMOS & \textit{r} & $2 \times 120$ & $21.29 \pm 0.06$ & \nod & $0.039$ \\
2018 Apr 18.452 & 0.17 & Gemini-N/GMOS & \textit{i} & $4 \times 120$ & $21.33 \pm 0.16$ & \nod & $0.029$ \\
2018 Apr 18.461 & 0.18 & Gemini-N/GMOS & \textit{g} & $4 \times 120$ & $22.07 \pm 0.15$ & \nod & $0.056$ \\
2018 Apr 18.469 & 0.19 & Gemini-N/GMOS & \textit{z} & $4 \times 120$ & $21.55 \pm 0.10$ & \nod & $0.022$ \\
2018 Apr 19.279 & 1.00 & Gemini-N/GMOS & \textit{r} & $15 \times 120$ & $23.86 \pm  0.13$ & \nod & $0.039$ \\
2018 Apr 19.398 & 1.12 & Gemini-N/GMOS & \textit{i} & $12 \times 120$ & $23.85 \pm 0.08$ & \nod & $0.029$ \\
2018 Apr 19.421 & 1.14 & Gemini-N/GMOS & \textit{z} & $12 \times 120$ & $23.37 \pm 0.29$ & \nod & $0.022$ \\
2018 Apr 19.458 & 1.18 & Gemini-N/GMOS & \textit{g} & $12 \times 120$ & $24.50 \pm 0.31$ & \nod & $0.056$ \\
2018 Apr 21.091 & 2.81 & Gemini-S/GMOS & \textit{r} & $18 \times 120$ & $24.95 \pm 0.13$ & \nod & $0.039$ \\
2018 Apr 23.073 & 4.79 & Gemini-S/GMOS & \textit{r} & $15 \times 180$ & $\approx$25.2$^{\star}$  & \nod & $0.039$ \\
2018 Apr 24.255 & 5.97 & UKIRT/WFCAM & \textit{J} & $63 \times 40$ & $\gtrsim 21.0$ & \nod & $0.012$ \\
2018 Apr 24.296 & 6.02 & UKIRT/WFCAM & \textit{K} & $63 \times 40$ & $\gtrsim 21.1$ & \nod & $0.005$ \\
2018 Apr 29.161 & 10.88 & MMT/MMIRS & \textit{J} & $29 \times 61.96$ & \nod & 23.34 $\pm$ 0.40 & $0.012$ \\ 
2018 Jun 7.757 & 50.48 & MMT/Binospec & \textit{r} & $13 \times 180$ & $\gtrsim 25.2$ & $25.50 \pm 0.43$ & $0.039$ \\
2018 Nov 19 & 215 & MMT/MMIRS & \textit{K} & $62 \times 30.98$ & \nod & $>22.0$ & $0.005$ \\
2018 Nov 27 & 223 & MMT/MMIRS & \textit{K} & $52 \times 30.98$ & \nod & $>22.4$ & $0.005$ \\
2019 Feb 1 & 289$^\dagger$ & Gemini-N/GMOS & \textit{r} & $14 \times 120$ & \nod & $25.69 \pm 0.21$ & $0.039$ \\
2019 Feb 1 & 289$^\dagger$ & Gemini-N/GMOS & \textit{i} & $16 \times 120$ & \nod & $24.82 \pm 0.14$ & $0.029$ \\
2019 Feb 1 & 289$^\dagger$ & Gemini-N/GMOS & \textit{z} & $20 \times 90$ & \nod & $24.62 \pm 0.21$ & $0.022$ \\
2019 Jun 18 & 426 & MMT/MMIRS & \textit{H} & $91 \times 30.98$ & \nod & $\gtrsim 22.8$ & $0.008$ \\
2020 Jan 10 & 632$^\dagger$ & MMT/MMIRS & \textit{J} & $29 \times 61.96$ & \nod & $\gtrsim 23.3$ & $0.012$ \\
2020 Mar 5 & 687 & MMT/MMIRS & \textit{Y} & $30 \times 119.49$ & \nod & $\gtrsim 23.3$ & $0.012$ \\
2020 Nov 20 & 947 & MMT/Binospec & \textit{g} & $20 \times 60$ & \nod & $\gtrsim 25.7$ & $0.056$ \\
\enddata
\tablecomments{$^{\star}$ While the HOTPANTS residual image for this epoch does not exhibit any source of meaningful significance, there is clearly afterglow flux contributing at this epoch based on differential photometry. The value reported here is thus based on differential photometry, assuming $r=25.69$~AB mag for the host galaxy. We do not, however, include this data point in our fitting.\\
$^{\dagger}$ These observations serve as template images to compute earlier afterglow fluxes. Limits correspond to $3\sigma$ confidence and uncertainties correspond to $1\sigma$. Magnitudes are corrected for Galactic extinction \citep{sf11}.}
\end{deluxetable*} 
%

\subsection{Optical and Near-infrared Observations}\label{sec:GRB180418A_opt_obs}

In addition to the {\it Swift}/UVOT detection of the optical afterglow of GRB\,180418A, there were several ground-based monitoring campaigns with optical facilities including: 25-cm T\'elescope \`a Action Rapide pour les Objets Transitoires (TAROT), RATIR mounted on the 1.5-m Harold L. Johnson Telescope \citep{Becerra2019}, 0.76-m Katzman Automatic Imaging Telescope \citep[GCN 22647;][]{ZhengW2018}, 2-m Faulkes Telescope North \citep[GCN 22648;][]{Guidorzi2018}, 1.5-m telescope at Observatorio de Sierra Nevada \citep[GCN 22657;][]{Sota2018}, 2.5-m Nordic Optical Telescope \citep[GCN 22660;][]{Malesani2018}, 2.2-m MPG telescope \citep[GCN 22662 and 22666;][]{Schady2018a,Schady2018b}, Xinglong 0.8-m Tsinghua-NAOC telescope \citep[GCN 22661;][]{XinL2018}, 3.6-m Devasthal Optical Telescope \citep[GCN 22663;][]{Misra2018}, 2.1-m Otto Struve telescope \citep[GCN 22668;][]{ChoiC2018} and Murikabushi 1-m telescope \citep[GCN 22670;][]{Horiuchi2018}. In the following section, we report on our optical afterglow and host galaxy imaging.

\subsubsection{Afterglow Imaging}\label{sec:GRB180418A_imaging}

\begin{figure*}
\includegraphics[width=\textwidth]{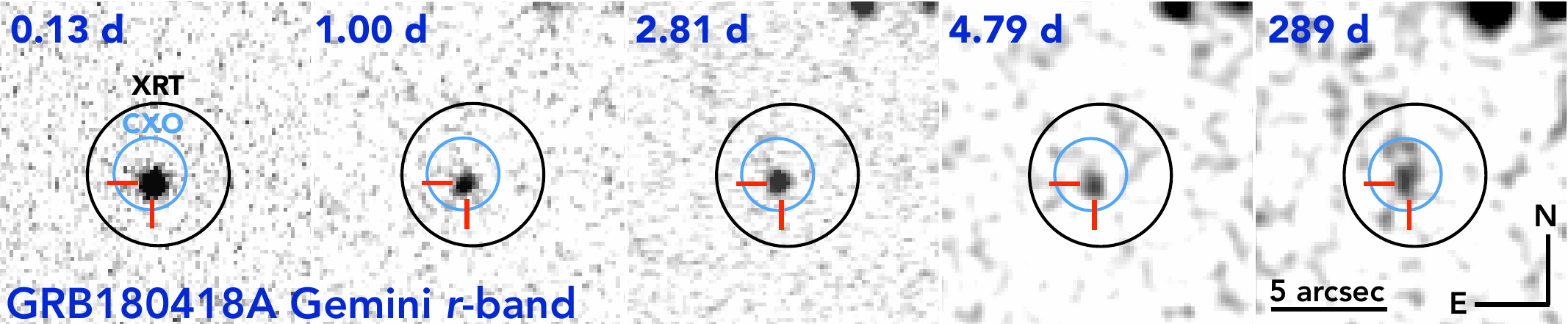}
\caption{Gemini-North and South GMOS $r$-band imaging sequence of the optical afterglow of \grb, over $\delta t = 0.13-4.79$~days. A deep, template image at $\delta t \approx 289$~days (last panel) reveals a faint, underlying host galaxy with $r=25.69 \pm 0.21$~mag. The position of the optical afterglow (red cross-hairs) is coincident with the \textit{Chandra} (CXO) position ($3\sigma$ radius including astrometric uncertainty; blue circle) and the enhanced XRT position ($3\sigma$ radius; black circle). The scale and orientation of the images are denoted in the last panel, and the last two panels have been smoothed for display purposes.}
\label{fig:optag}
\end{figure*}

We triggered Target-of-Opportunity (ToO) observations of the location of \grb\ with the Gemini Multi-Object Spectrograph (GMOS; Program GN-2018A-Q-121) mounted on the 8-m Gemini-North telescope on 2018 Apr 18 UT starting at $\delta t = 3.1$~hr. We obtained observations in the $griz$-bands, and used standard tasks in the IRAF/{\tt gemini} package to create bias- and flat-field frames, apply them to the science images, and co-add the images in each filter. On the outskirts of the enhanced XRT position, we clearly detect an optical point source coincident with the {\it Chandra} X-ray position in all bands (Figure~\ref{fig:optag}). The details of our observations are listed in Table~\ref{tab:gb_obs}.

To track the fading and color evolution of the source, we obtained an additional set of $griz$-band observations with the GMOS instruments mounted on the 8-m Gemini-North and Gemini-South telescopes on 2018 Apr~19 UT starting at $\delta t = 24.0$~hr, as well as $r$-band observations at two additional epochs of $\delta t = 2.89$~days and $4.79$~days. The last of these observations still clearly exhibits a detected source (Figure~\ref{fig:optag}), necessitating late-time, deeper observations to assess the contribution from the underlying host galaxy (see Section~\ref{sec:GRB180418A_hostobs}). Therefore, we obtained $riz$-band observations of the field with Gemini-North/GMOS at $\delta t \approx 289$~days (Program GN-2018B-Q-117), which have significantly greater depth than the previous epochs and thus serve as adequate template images for the previous imaging. For each filter, we perform image subtraction between each of the earlier epochs and the late-time observation with the {\tt HOTPANTS} software package \citep{bec15}. 

Calibrated to SDSS DR12, we use {\tt SExtractor} to derive an optical afterglow position of RA=\ra{11}{20}{29.20} and Dec=$+$\dec{24}{55}{58.83} (J2000) with a $1\sigma$ positional uncertainty of $0.12''$, including the contributions from the afterglow centroid and the astrometric tie uncertainty to SDSS. This position is fully consistent with the {\it Chandra} afterglow position (Figure~\ref{fig:optag}). We perform aperture photometry on the residual images with the IRAF/{\tt phot} package, using an aperture of 2.5$\times \theta_{\rm FWHM}$ for each epoch and filter. The $r$-band afterglow observations are displayed in Figure~\ref{fig:optag} and the resulting photometry is listed in Table~\ref{tab:gb_obs}. We note that the data at $\delta t \approx 4.79$~days is based on differential photometry, and we do not include this point in subsequent fitting.

To place limits on any transient emission on timescales of $\gtrsim$\,few days, we also obtained near-infrared (NIR) imaging in the $J$ and $K$-bands with the Wide-field Camera (WFCAM; \citealt{caa+07}) on the 3.8-m United Kingdom Infrared Telescope (UKIRT) at $\delta t\approx 6.0$~days. We obtained pre-processed images from the WFCAM Science Archive \citep{hcc+08} which are corrected for bias, flat-field, and dark current by the Cambridge Astronomical Survey Unit\footnote{http://casu.ast.cam.ac.uk/}. For each epoch and filter, we co-add the images and perform astrometry relative to 2MASS using a combination of tasks in Starlink\footnote{http://starlink.eao.hawaii.edu/starlink} and IRAF. We do not detect any emission coincident with the afterglow, and measure upper limits of $J \gtrsim 21.0$~mag and $K \gtrsim 21.1$~mag (calibrated to 2MASS and converted to the AB system) based on faint sources in the vicinity of the GRB position.

Finally, we obtained $J$-band observations with Magellan Infrared Spectrograph (MMIRS) and $r$-band observations with Binospec, both mounted on the 6.5-m MMT (Multiple Mirror Telescope) at $\delta t \approx 10.9$~days and $50.5$~days, respectively. We used custom pipelines\footnote{\url{https://github.com/CIERA-Transients/Imaging_pipelines/blob/master/MMIRS_pipeline.py}, \url{https://github.com/CIERA-Transients/Imaging_pipelines/blob/master/BINOSPEC_pipeline.py}} using routines from \texttt{ccdproc} \citep{Craig2017} and \texttt{astropy} (\citealt{astropy:2013}, \citealt{astropy:2018}) to perform bias subtraction, flat-fielding and gain correction calibrations, as well as additional sky subtraction routines for MMIRS to take into account the varying IR sky. We aligned and co-added the data, and calibrated to 2MASS and SDSS, respectively. Performing image subtraction with {\tt HOTPANTS} relative to later template images (see Section~\ref{sec:GRB180418A_hostobs}), we place limits on late-time transient emission of $r \gtrsim 25.2$~mag (Table~\ref{tab:gb_obs}).

We briefly compare our limits to the luminosities of GRB-SNe. In particular, we compare the final $r$-band upper limit at $\delta t \approx 50$~days to the optical emission of GRB-SN1998bw, associated with the long GRB\,980425 \citep{Galama1998}. At an assumed $z=1$ (see Section\,\ref{sec:GRB180418A_hostobs}), we find that our upper limit of $\nu{\rm L}_{\nu}\gtrsim7.8 \times 10^{42}$~erg~s$^{-1}$ is not deep enough to constrain the presence of a supernova as luminous as SN1998bw ($\approx 10^{42}$\,erg~s$^{-1}$; \citealt{Clocchiatti2011}) in the appropriate rest-frame band and time ($U$-band and $\delta{\rm t_{rest}} \approx 25.2$\,days at $z=1$). Only if GRB\,180418A originated at lower redshifts of $z<0.5$ could we effectively use this limit to rule out the existence of SN1998bw-like emission.

%

\subsubsection{Afterglow Spectroscopy}\label{sec:GRB180418A_Spectroscopy}

Using the Gemini-North rapid ToO program, we obtained a set of $4 \times 900$~s of spectroscopy of the optical afterglow (initially reported in \citealt{GCN22659}) on 2018 Apr 18 UT at a mid-time of $\delta t = 2.4$~hr. We obtained a pair of exposures with the R400 grating at each of two central wavelengths, 5200 \AA\ and 5250 \AA, covering a wavelength range of $4500-7600$ \AA. We used the Gemini IRAF package to apply bias and flat-field corrections, cosmic ray rejection, and to align and stack the frames. We additionally used CuAr lamp spectra for wavelength calibration that were taken during the observations, and a spectrum of standard star HZ44 taken on 2018 February 28 with the same setup to obtain a relative flux calibration. The resulting spectrum exhibits a featureless blue continuum, with no notable features in emission or absorption that could be attributed to the host galaxy. We note that the faintness of the host galaxy (Section~\ref{sec:GRB180418A_hostobs}) precludes a strong statement on the presence of emission features, but overall exhibits no strong nebular emission.

\subsubsection{Host Galaxy Observations}
\label{sec:GRB180418A_hostobs}

In Gemini imaging at $\delta t \approx 289$~days, we identify a faint galaxy at RA=\ra{11}{20}{29.21} and Dec=$+$\dec{24}{55}{58.73} (J2000), coincident with the {\it Chandra} and Gemini afterglow positions (Figure~\ref{fig:optag}). We perform aperture photometry using the IRAF/{\tt phot} package as previously described and measure a brightness of $r = 25.69 \pm 0.21$~mag. The galaxy is at an angular offset from the optical afterglow position of $0.16 \pm 0.04''$. Using this offset and the $r$-band magnitude (Table~\ref{tab:gb_obs}), we calculate the probability of chance coincidence following the methods of \citet{bkd02} to be $P_{cc}=1.4 \times 10^{-3}$. The low value of $P_{cc}$, coupled with the fact that there are no other detectable $>3\sigma$ sources within $7.5''$ of the afterglow position to $r \gtrsim 26$~mag, solidifies this source as the host galaxy. The host galaxy is detected in $riz$-bands, and the photometry results are in Table~\ref{tab:gb_obs}. We also obtain a deep upper limit with MMT/Binospec observations in $g$-band of $g \gtrsim 25.7$~mag.

We additionally obtained NIR imaging observations in the $YJHK$-bands with the MMT/MMIRS. Only the $J$-band image yields a host galaxy detection of $J=$\,23.34$\pm$\,0.40~mag. For the remaining filters, we calculate $3\sigma$ upper limits based on faint sources in the vicinity of the GRB in each image. The measurements and $3\sigma$ upper limits for the remaining filters are listed in Table~\ref{tab:gb_obs}.

%
\begin{figure}
	\centering
	\includegraphics[width=\columnwidth]{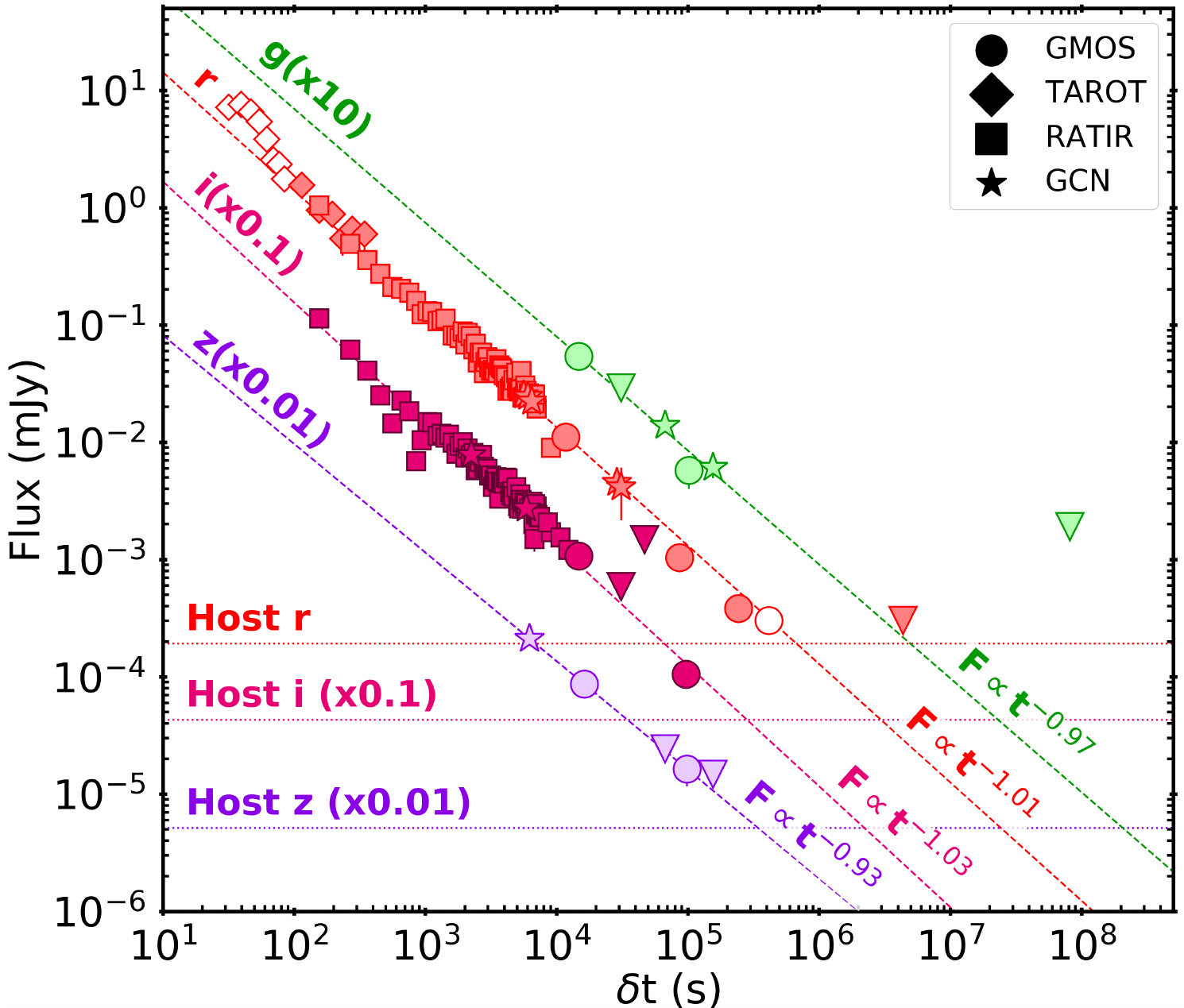}
    \caption{Optical afterglow light curves of GRB\,180418A in the $griz$ filters. Circles represent our new Gemini afterglow data (Table~\ref{tab:gb_obs}). Literature data from TAROT (diamonds), RATIR (squares) and other sources (stars) are also shown \citep{Becerra2019,ChoiC2018,Guidorzi2018,Horiuchi2018,Malesani2018,Misra2018,Schady2018a,Schady2018b,Sota2018,XinL2018}. Triangles indicate $3\sigma$ upper limits. Only those observations in the $riz$ filters for which the host galaxy contribution is less than 5$\%$ of the total optical flux are plotted. Observations in the $r$-band which are ignored in the power-law fit are shown as open symbols. Horizontal lines denote the flux of the host galaxy, while dashed lines indicate the best-fit power-law decay models for the different bands.}
    \label{fig:180418a_OptIR_lc}
\end{figure}
%

\section{Analysis \& Results}\label{sec:GRB180418A_analysis}

\subsection{Redshift Estimate}\label{sec:GRB180418A_redshift}

To estimate the redshift of \grb, we consider both the detection of the afterglow and the inferred luminosity of the host galaxy. The detection of the afterglow in the $uvm2$ UVOT filter (\citealt{Siegel2018}), with $\lambda_{\rm max} \approx 2964$\,\AA\ (the wavelength at the upper end of the bandpass), automatically places an upper limit on the redshift of $z<2.25$, corresponding to the Lyman limit of $\lambda=912$\AA\ at these redshifts, as a higher redshift would result in the complete suppression of flux at these wavelengths. On the other hand, the featureless afterglow spectrum implies that $z \gtrsim 1$ (or that the burst sightline did not intersect with any strong absorption features). Moreover, if GRB\,180418A originated at the median redshift of short GRBs of $z=0.5$, the inferred host luminosity would be low, with $L \lesssim 0.01L^{*}$, where we expect only $\approx5\%$ of the stellar mass at z=0.5 to reside in galaxies fainter than this, implying this is unlikely \citep{Tomczak2014}. Thus, we constrain a most likely redshift range of $z \approx$\,1-2.25 for GRB\,180418A. This is also in agreement with the results of \cite{Becerra2019}, who found $z \approx 0.3-1.31$ based on the photometric upper limits and the combined X-ray, UV, and optical broadband spectral energy distribution.

\subsection{Afterglow of GRB\,180418A}\label{sec:GRB180418A_afterglow}

\subsubsection{Light curve Fitting and Spectral Parameters Determination}\label{sec:GRB180418A_LC_fitting}

To quantify the temporal evolution of the afterglow of GRB\,180418A and the spectral information in the X-ray and optical bands, we consider the general relation $F_\nu \propto \text{t}^{\alpha} \nu^{\beta}$, where $\alpha$ and $\beta$ are the temporal and spectral power-law indices respectively. In particular, we determine $\alpha_\text{X}$ and $\alpha_\text{opt}$ by fitting the light curves in each band with a single power-law model, $F_\nu \propto \text{t}^{\alpha}$, using a $\chi^2$-minimization procedure with a best-fit normalization defined by:

\begin{equation}
    \text{C} = \frac{\sum_{\text{i}=1}^{\text{N}}\frac{F_{\text{model, i}} \times F_{\nu, \text{i}}}{\sigma^2_{\nu, \text{i}}}}{\sum_{\text{i}=1}^{\text{N}}\frac{F_{\text{model, i}}^2}{\sigma^2_{\nu, \text{i}}}}
\end{equation}

\noindent where $F_\text{model, i}$ and $F_{\nu, \text{i}}$ are the un-normalized and observed fluxes respectively, $\sigma_{\nu, \text{i}}$ are the uncertainties on the fluxes, and N is the number of data points.

To fit the X-ray light curve, we include all data points (see Figure~\ref{fig:180418a_Xray_lc}). For the optical afterglow, since we are only interested in the forward shock (FS) afterglow behavior, we ignore data at $\delta t \leq 100$\,s in the $r$-band light curve as there is an initial flux density enhancement that has been attributed to a reverse shock (Figure~\ref{fig:180418a_OptIR_lc}; \citealt{Becerra2019} and Section\,\ref{sec:GRB180418_RSFS_scenario}), and include all available data in the $giz$-bands. The final temporal indices we obtain are $\alpha_{\text{X}} = -0.98_{-0.16}^{+0.12}$ and $\langle \alpha_{\text{opt}} \rangle = -1.01 \pm 0.03$ with 1$\sigma$ uncertainties, where $\langle \alpha_{\text{opt}} \rangle$ is the weighted mean of the temporal indices corresponding to the four optical bands: $\alpha_{\text{g}} = -0.97 \pm 0.13$, $\alpha_{\text{r}} = -1.01_{-0.04}^{+0.03}$, $\alpha_{\text{i}} = -1.03 \pm 0.04$ and $\alpha_{\text{z}} = -0.93_{-0.16}^{+0.14}$.

We determine the X-ray spectral index, $\beta_{\text{X}}$, from the relation $\beta_{\text{X}} \equiv 1 - \Gamma_{\text{X}}$, where $\Gamma_{\text{X}}$ is the X-ray spectral photon index. We calculate the value of $\beta_{\text{X}}$ for each X-ray observation using the $\Gamma_{\text{X}}$ spectral values and obtain the weighted mean of $\langle \beta_{\text{X}} \rangle = -0.85 \pm 0.14$ (1$\sigma$ uncertainty). In the case of the optical band, we utilize contemporaneous observations at $\delta t \approx 0.13-0.19$~days in the Gemini $griz$-bands, and extrapolate them to a common time of $\delta t \approx 0.13$~days to determine $\beta_{\text{opt}}$. We use $\chi^2$-minimization to fit a single power-law, finding $\beta_{\text{opt}} = -0.70 \pm 0.19$ (1$\sigma$ uncertainty).

\subsubsection{Energy and Circumburst Density Properties}\label{sec:GRB180418A_energy_density}

In this section, we model the detected emission from \grb\ in the different bands (optical and X-rays) as well as limits (NIR and radio) in the framework of the standard synchrotron forward shock model. In this scenario, the broad-band emission originates from a forward shock resulting from the interaction of the relativistic GRB jet with the surrounding environment \citep{Sari1998,Granot2002}. The model is defined by the following parameters: isotropic-equivalent energy of the jet (E$_{\rm K, iso}$), circumburst density ($n$), power-law index of accelerated electrons ($p$), fractions of the post-shock energy transmitted to electrons ($\epsilon_{\rm e}$) and magnetic field ($\epsilon_{\rm B}$), and the opening angle of the jet ($\theta_{j}$). Likewise, the synchrotron spectral shape is characterized by the synchrotron self-absorption frequency ($\nu_{sa}$), the synchrotron peak frequency ($\nu_{m}$) and the cooling frequency ($\nu_{c}$).

First, we determine the position of the X-ray band with respect to the cooling frequency, assuming a constant density medium. For that, we calculate the value of $p$ using the relations between the temporal and spectral indices introduced by \citet{Granot2002}. We require the value of $p$ to be consistent within the errors in one of the next two scenarios: $\nu_{\rm X} > \nu_\text{c}$ or $\nu_{\rm{m}} < \nu_{\rm X} < \nu_\text{c}$. For $\nu_{\rm X} > \nu_\text{c}$, we find that the values of $p$ are inconsistent and, furthermore, lead to $p < 2$. This is an unlikely value since $p$ generally ranges between 2 and 3 as a direct consequence of the Lorentz factor distribution \citep[e.g.,][]{Jager1992}. On the other hand, for $\nu_{\text{m}} < \nu_{\rm X} < \nu_\text{c}$, we obtain consistent values of $p$ within the errors for both X-ray and optical bands. Therefore, we accept the scenario where $\nu_{\text{m}} < \nu_{\rm opt} < \nu_{\rm X} < \nu_\text{c}$ for the duration of the observations, and calculate a weighted mean value of $\langle p \rangle = 2.39 \pm 0.12$.

In addition, given the borderline nature of the classification of GRB\,180418A (short vs. long), we briefly explore the possibility that the shock-wave expands into a wind medium, with $n(r) \propto r^{-2}$, as expected for massive star progenitors. We follow the consequent closure relations for a wind environment from \cite{Granot2002}, but find inconsistent values of $p$, as well as $p<2$ for both of the aforementioned scenarios. Given that a large fraction of bona fide long GRB afterglows are inconsistent with the wind medium solution \citep{Racusin2009,Schulze2011,Laskar2018a}, we note that this alone is not conclusive as to the nature of the progenitor for GRB\,180418A. For our subsequent analysis, we consider a constant-density interstellar medium (ISM).

Next, we constrain the physical burst properties, $E_{\rm {K, iso}}$ and $n$, utilizing the data in the X-ray and optical bands, and the radio upper limit. Specifically, we use $F_{\rm {\nu,X}} = (2.48 \pm 0.23)\times 10^{-4}$\,mJy at $\delta t \approx 0.05$ days and $\nu_{\rm {X}}=4.19 \times 10^{17}$\,Hz (log-centered frequency of the 0.3$-$10\,keV energy band), $F_{\rm {\nu, opt}} = (1.10 \pm 0.12)\times10^{-2}$\,mJy at $\delta t\approx 0.13$ days and $\nu_{\rm {opt}}=4.84 \times 10^{14}$\,Hz, and $F_{\nu,{\rm {radio}}}< 9.9 \times 10^{-2}$\,mJy at $\delta t \approx 0.61 $ days and $\nu_{\rm {radio}}=15.5 \times 10^{9}$\,Hz. We calculate the $E_{\rm {K,iso}}$-$n$ relations set by the broad-band observations fixing the value of $\epsilon_{\rm {e}}$ to 0.1 \citep{Panaitescu2002,Sironi2011} and varying the value of $\epsilon_{\rm {B}}$ between $10^{-4}-0.1$ considering $z=1$ and $z=1.5$. Assuming $\nu_{\rm{sa}} < \nu_{\rm {radio}} < \nu_{\rm{m}}$ and $\nu_{\rm{m}} < \nu_{\rm {opt}} < \nu_{\rm{X}} < \nu_{\rm{c}}$, we set the minimum value of the cooling frequency ($\nu_{\rm{c, min}}=2.4 \times 10^{18}$\,Hz) equal to the upper edge of the X-ray band (equivalent to 10\,keV), which translates to an upper limit on the $E_{\rm {K,iso}}$-$n$ parameter space. Finally, we set the minimum value of the circumburst density to $n_{\rm {0, min}}=10^{-4}$~cm$^{-3}$, determined by the low end of typical ISM particle densities. 

Combining the probability distributions of $E_{\rm K,iso}$ and $n$, and assuming values of $\epsilon_{\rm{B}}$ ranging between $0.1-10^{-4}$, we find that $E_{\rm{K, iso}}=(0.89-29) \times 10^{52}$\,erg and $n=(2.56-56) \times 10^{-4}$\,cm$^{-3}$ at $z=1$, whereas E$_{\rm{K, iso}}=(1.64-35) \times 10^{52}$\,erg and $n=(2.21-160)\times 10^{-4}$\,cm$^{-3}$ at $z=1.5$. Lastly, we use these values of $E_{\rm{K, iso}}$ and $E_{\rm{\gamma, iso}}$ (Section~\ref{sec:GRB180418A_classification}) to calculate the $\gamma$-ray efficiency of $\eta \approx 0.2-0.01$ at $z=1$ and $\eta \approx 0.1-0.01$ at $z=1.5$. The results are listed in Table~\ref{tab:GRB180418A_energy_density}.

%
\begin{deluxetable*}{ccccccccc}[!t]
\tabletypesize{\normalsize}
\tablecolumns{9}
\tablewidth{0pc}
\tablecaption{GRB\,180418A burst properties and circumburst density\label{tab:GRB180418A_energy_density}}
\tablehead{
\colhead{Case} & \colhead{$\epsilon_{\rm{B}}$} & \colhead{E$_{\rm{K,iso}}$} & \colhead{$n_0$} & \colhead{$\langle \theta_{\rm j, min} \rangle$} & \colhead{$f_b$} & \colhead{E$_{\rm{K, lim}}$} & \colhead{E$_{\gamma\rm{, lim}}$} & \colhead{$\eta$}\\
& & \colhead{(erg)} & \colhead{(cm$^{-3}$)} & \colhead{($^{\circ}$)} & & \colhead{(erg)} & \colhead{(erg)} &
}
\startdata
\hline\hline
\multicolumn{9}{c}{z=1} \\
\hline\hline
Case A & 0.1 & $(8.9^{+1.1}_{-1.0}) \times 10^{51}$ & $(2.56^{+0.88}_{-0.66}) \times 10^{-4}$ & $10.45^{+0.52}_{-0.53}$ & 0.017 & $1.51 \times 10^{50}$ & $4.61 \times 10^{49}$ & 0.23\\
Case B & $10^{-2}$ & $(2.07^{+1.22}_{-0.77}) \times 10^{52}$ & $(1.7^{+4.2}_{-1.2}) \times 10^{-3}$ & $11.9^{+2.9}_{-2.3}$ & 0.022 & $4.55 \times 10^{50}$ & $5.96 \times 10^{49}$ & 0.12\\
Case C & $10^{-3}$ & $(4.8^{+6.0}_{-2.7}) \times 10^{52}$ & $(1.1^{+9.1}_{-1.0}) \times 10^{-2}$ & $13.6^{+6.2}_{-4.2}$ & 0.028 & $1.34 \times 10^{51}$ & $7.59 \times 10^{49}$ & 0.05\\
Case D & $10^{-4}$ & $(2.9^{+1.9}_{-1.2})\times10^{53}$ & $(5.6^{+16.2}_{-4.1}) \times 10^{-3}$ & $9.9^{+2.6}_{-2.1}$ & 0.015 & $4.4 \times 10^{51}$ & $4.1 \times 10^{49}$ & 0.009\\
\hline\hline
\multicolumn{9}{c}{z=1.5} \\
\hline\hline
Case E & 0.1 & $(1.64 \pm 0.10) \times 10^{52}$ & $(2.21^{+0.17}_{-0.16}) \times 10^{-4}$ & $8.74^{+0.13}_{-0.12}$ & 0.012 & $1.97 \times 10^{50}$ & $7.14 \times 10^{49}$ & 0.14\\
Case F & $10^{-2}$ & $(3.8^{+1.8}_{-1.2}) \times 10^{52}$ & $(1.47^{+2.61}_{-0.94}) \times 10^{-3}$ & $9.9^{+1.9}_{-1.6}$ & 0.015 & $5.7 \times 10^{50}$ & $8.9 \times 10^{49}$ & 0.07\\
Case G & $10^{-3}$ & $(8.9^{+9.6}_{-4.6}) \times 10^{52}$ & $(9.8^{+60.5}_{-8.4}) \times 10^{-3}$ & $11.4^{+4.5}_{-3.2}$ & 0.020 & $1.78 \times 10^{51}$ & $1.19 \times 10^{50}$ & 0.03\\
Case H & $10^{-4}$ & $(3.5^{+3.7}_{-1.8}) \times 10^{53}$ & $(1.6^{+9.2}_{-1.3}) \times 10^{-2}$ & $10.1^{+4.0}_{-2.9}$ & 0.016 & $5.6 \times 10^{51}$ & $9.5 \times 10^{49}$ & 0.008\\
\enddata
\tablecomments{The median values of the isotropic-kinetic energy (E$_{\rm{K,iso}}$) and circumburst density (n$_0$) for values of $\epsilon_{\rm B}= 0.1-10^{-4}$ at $z=1$ and $z=1.5$ (Cases A-D and Cases E-H, respectively). The median values of the minimum opening angle of the jet (Section\,\ref{sec:GRB180418A_angle}) for each case are represented by $\langle \theta_{\rm j, min} \rangle$. The parameter $f_{\rm b}$ is the beaming factor and E$_{\rm K, lim}$ corresponds to the lower limit of the true $\gamma$-ray and kinetic energy values. The $\eta$ parameter is defined as E$_{\rm{\gamma,iso}}$/(E$_{\rm{K,iso}}$+E$_{\rm{\gamma,iso}}$). Errors are 1$\sigma$.}
\end{deluxetable*}
%

\subsubsection{Constraints on the Jet Opening Angle}\label{sec:GRB180418A_angle}

Here, we study our late-time monitoring of the X-ray afterglow to determine the jet opening angle ($\theta_{\rm j}$) of \grb. In the fireball model, the observed temporal behavior from a spherical expansion for an on-axis observer is initially similar to that of a collimated relativistic outflow \citep{Rees1992,Meszaros1992a,Meszaros1993,Sari1995}. As the value of the bulk Lorentz factor ($\Gamma$) declines over time to reach a value of $\theta_{\rm j}^{-1}$ \citep{Piran1995}, a significant temporal steepening in the afterglow light curve is expected for a collimated outflow, known as a `jet break' \citep{Sari1999,vanEerten2013}, after which the flow may undergo lateral expansion \citep{Granot2012}. From the detection of the jet break in the afterglow light curve at a certain time, one can derive the GRB jet opening angle. In contrast, a spherical outflow is expected to decline as a single power-law until it reaches the non-relativistic regime \citep{Taylor1950,Sedov1959,vanEerten2010b,Sironi2013}.

For \grb, the optical afterglow in the $griz$-bands exhibits a single power-law decline to $\delta t \approx 2.8$~days. In the X-ray band, the afterglow light curve of \grb\ is well-modeled as a single power-law decay up to $\delta t \approx 38.5$ days and does not show any noticeable deviation from this decline rate. Thus, we can determine a lower limit for $\theta_{\rm j}$ by using the time of the last \textit{Chandra} observation ($\delta t = 38.515$~days), following the relation given by \citet{Sari1999} and \citet{Frail2001},

\begin{equation}
    \theta_{\rm j}\,\geq\,37.53\,(1+z)^{-3/8}\,E_{\text{K, iso, 52}}^{-1/8}\,n_{0}^{1/8}~~~{\rm [deg]}
\label{eqn:jb}
\end{equation}

\noindent where E$_{\rm{K, iso, 52}}$ is in units of $10^{52}$\,ergs and n$_{0}$ is in units of cm$^{-3}$. We calculate the minimum value, $\theta_{\rm j,min}$, using Equation~\ref{eqn:jb} for every allowed pair of E$_{\rm{K,iso}}-n$ as determined in Section~\ref{sec:GRB180418A_energy_density}, and compute the resulting cumulative probability distribution for each value (Figure~\ref{fig:180418a_OpeningAngle}). The median values of the minimum opening angles, $\langle \theta_{\rm j, min} \rangle$, are listed in Table~\ref{tab:GRB180418A_energy_density}. Given that we do not detect a jet break in the afterglow light curve, we determine lower limits of $\langle \theta_{\rm j, min} \rangle = 9.9-13.6^{\circ}$ for $z=1$, and $\langle \theta_{\rm j, min} \rangle = 8.74-11.4^{\circ}$ for $z=1.5$. This is in agreement with the result reported in \citealt{Becerra2019}, where the jet opening angle is constrained to $\theta_{\rm j}\ge 7^{\circ}$ considering $z=0.5$ and the multi-wavelength information up to $0.8$\,days. We further note that if higher values for the density ($\approx 0.1$~cm$^{-3}$) and/or lower values of $\epsilon_B\lesssim 10^{-4}$ are considered, as suggested by the multi-wavelength modeling, then we obtain a wider opening angle constraint of $\theta_{\rm j, min} \approx 17^{\circ}$ for $z=1$.

Finally, we calculate the beaming correction factor (defined as $f_{\rm{b}} \equiv [1- \rm{cos}(\theta_{\rm{j}})]$). For every value of $\langle \theta_{\rm j, min} \rangle$, we obtain lower limits on the true kinetic energy, E$_{\rm{K}}=f_{\rm{b}}$\,E$_{\rm{K, iso}}$, as a wider jet would indicate a value closer to the isotropic-equivalent value. For the different values of $\epsilon_{\rm B}$ considered in this work, we obtain E$_{\rm{K,lim}}=(1.51-44)\times 10^{50}$\,erg at $z=1$, and E$_{\rm{K,lim}}=(1.97-56)\times 10^{50}$\,erg at $z=1.5$ (see Table~\ref{tab:GRB180418A_energy_density}).

\subsection{Reverse Shock Scenario}\label{sec:GRB180418_RSFS_scenario}

%
\begin{figure*}
	\centering
	$\vcenter{\hbox{\includegraphics[width=0.45\textwidth]{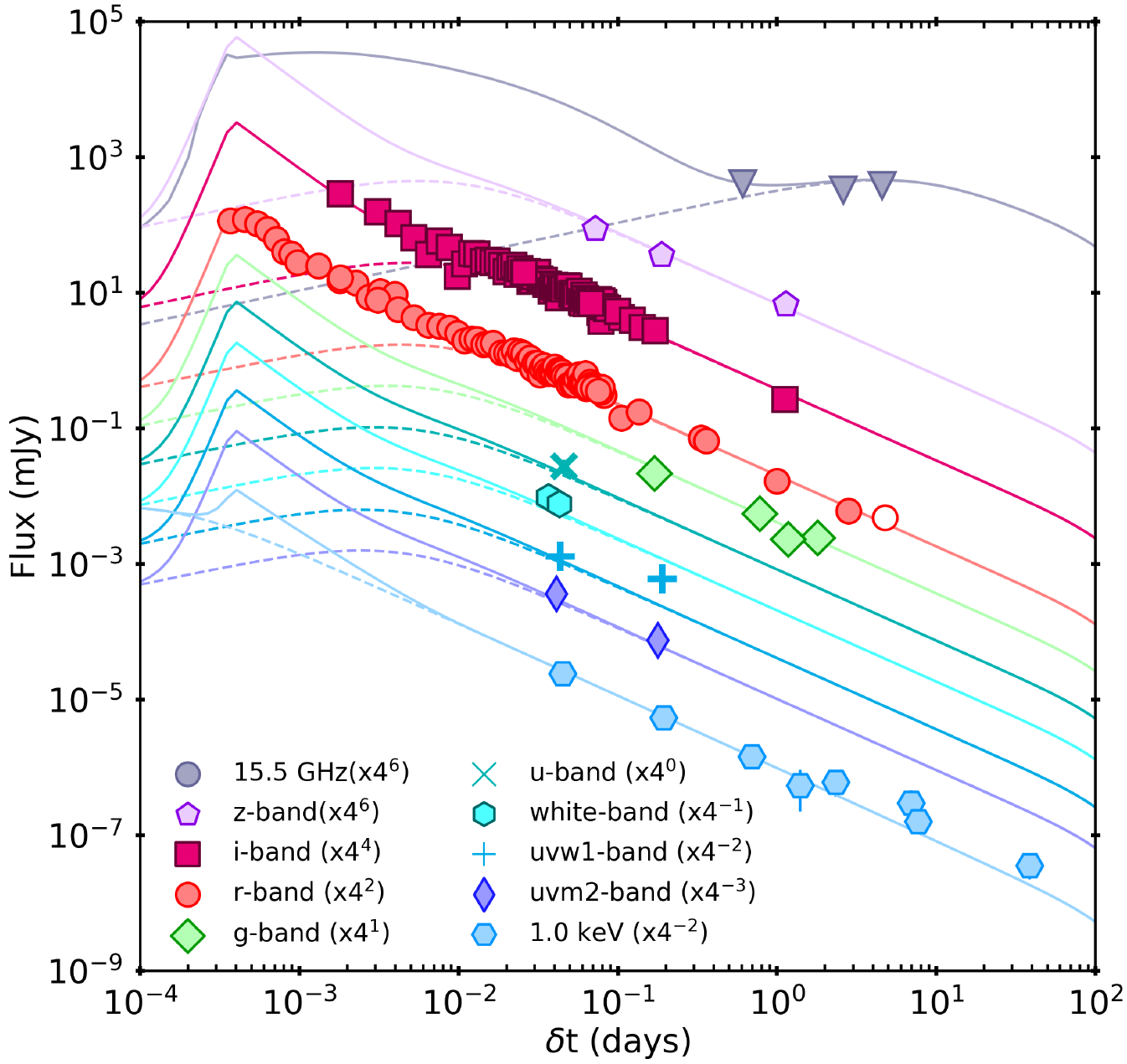}}}$
\hspace*{.1in}
$\vcenter{\hbox{\includegraphics[width=0.5\textwidth]{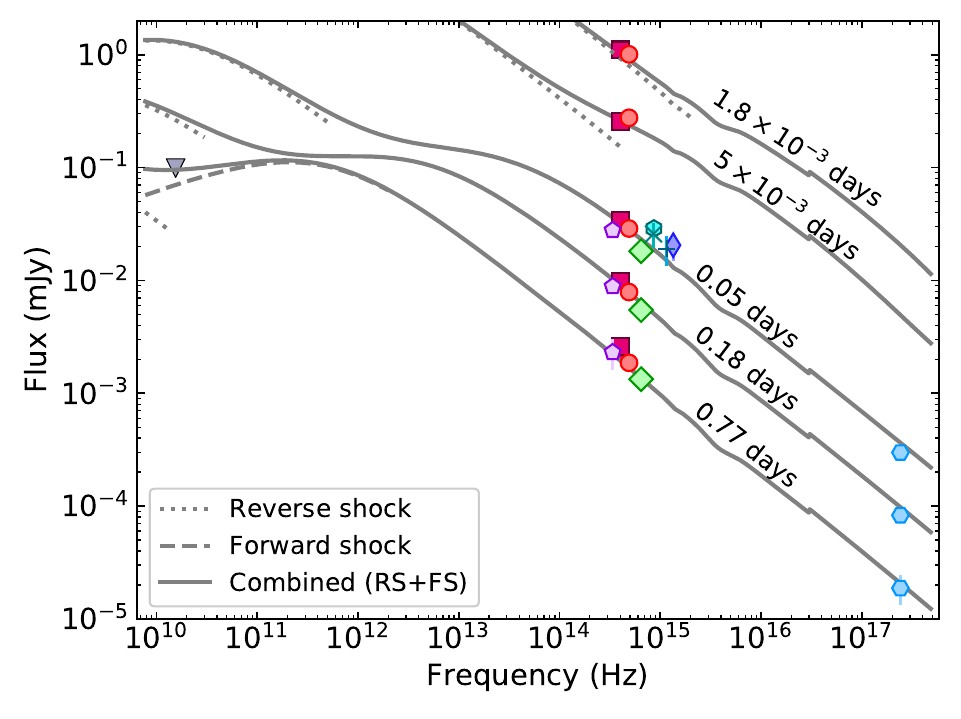}}}$
    \caption{{\it Left:} The radio to X-ray light curves of the GRB\,180418A afterglow and the best-fit RS+FS model (solid lines) with a magnetization parameter of $R_{\rm B}\approx5.2$ and an initial jet Lorentz factor of $\Gamma_0\approx150$. For each band, the FS component is indicated with dashed lines. For completion, we have considered the UVOT data (converted to AB system and corrected from Galactic extinction) reported by \cite{Siegel2018} in our modeling. Open symbols indicate data that are not included in the fitting (see Section\,\ref{sec:GRB180418A_imaging}). {\it Right:} Radio to X-ray spectral energy distribution of the GRB\,180418A afterglow spanning $1.8\times10^{-3}$~days to 0.77~days after the burst, together with the best-fit model (solid lines) decomposed into reverse (dotted) and forward shock (dashed) components. The radio upper limit constrains the peak flux and frequency of the FS spectrum, necessitating an RS component in the optical before $\approx0.05$~days.}
    \label{fig:180418a_multimodel}
\end{figure*}

Here, we explore the broad-band emission of GRB\,180418A in the context of a combined FS and reverse shock (RS) model, the latter of which propagates back into the ejecta, decelerating it \citep{Sari1998,Zhang2005}. This is in part motivated by the results of \citet{Becerra2019}, which explained the $r$-band afterglow at $\delta t \lesssim 10^{-3}$~days with an RS model (Figure\,\ref{fig:180418a_multimodel}). As the early optical and radio observations are the most relevant for this component, we first consider the locations of $\nu_{\rm radio}$ and $\nu_{\rm opt}$ with respect to $\nu_m$ at early times. At the time of the first radio upper limit, $\delta t \approx 0.61$~days, we calculate a limit on the radio-to-optical spectral index of $\beta_{\rm radio-opt}\gtrsim-0.36$, which is shallow compared to $\beta_{\rm opt} \approx -0.7$ (Section~\ref{sec:GRB180418A_LC_fitting}). This indicates that $\nu_{\rm radio} < \nu_{\rm m} < \nu_{\rm opt}$ at this time; this constraint allows us to derive limits on the peak frequency and flux of the forward shock of $\nu_{\rm m}\gtrsim4.7\times10^{11}$~Hz and $F_{\nu,\rm FS,{\rm max}}\lesssim0.3$~mJy, respectively.

In the constant density environment considered here, and which is also favored by the shallow optical and X-ray decay at $\delta t\gtrsim0.02$~days (Figure\,\ref{fig:180418a_multimodel}), the peak flux of the spectrum remains constant as $\nu_{\rm m}$ cascades to lower frequencies due to adiabatic cooling. However, the observed $r$-band flux remains greater than $F_{\nu,\rm FS,max}$ at early times, peaking at $\approx7$~mJy, or at least a factor of $\approx23$ brighter (Figure\,\ref{fig:180418a_multimodel}). Furthermore, the limit on $\nu_{\rm m,FS}$ implies that $\nu_{\rm m,FS}$ passes through $r$-band at $\delta t\gtrsim5\times10^{-3}$~days. Thus, the $r$-band emission at early times is too bright to be explained solely by FS emission.

One possible mechanism that can produce radiation in excess of the FS emission in the optical bands at early times is an RS. Early excess optical emission has been ascribed to a RS component for many long-duration GRBs \citep[][]{Laskar2018a,Laskar2019}, as well as for two short-duration GRBs, 051221A and 160821B  \citep{Soderberg2006,Lloyd2018,Lamb2019}, and possibly GRB\,200522A \citep{Fong2020}. Similar to that for the FS, the RS synchrotron spectrum is also characterized by an injection break ($\nu_{\rm m,RS}$) and cooling break ($\nu_{\rm c,RS}$), as well as a self-absorption break ($\nu_{\rm a,RS}$), although the latter cannot be constrained by our present data. RS emission is expected to peak at the deceleration time, $t_{\rm dec}$, when the RS reaches the back of the jet. The subsequent light curves depend on the hydrodynamics of the reverse-shocked shell, which, for short-duration GRBs, are expected to follow the thin-shell regime ($t_{\rm dec}\gtrsim {\rm T}_{90}$), resulting in a Newtonian RS \citep{kob00}. In this regime, the post-shock bulk Lorentz factor evolves with radius as $\Gamma\propto R^{-g}$, where $g$ is $\approx2.2$ in a uniform-density external environment \citep{ks00}. Considering $g\approx2.2$ and $p\approx2.4$ (as inferred for the FS), and the equations for $F_{\rm m,RS}$ at $\nu<\nu_{\rm m,RS}$ and $\nu>\nu_{\rm m,RS}$ introduced by \cite{ks00} in their Section~3.3, we expect $\alpha\approx-0.46$ before the passage of $\nu_{\rm m,RS}$ and $\alpha\approx-2.0$ thereafter. The observed $\alpha_{\rm r}$ of -1.38$\pm0.03$ lies between these expected limits at $(0.4-5)\times10^{-3}$~days. 

One explanation may be that for the observed $r$-band light curve at $\lesssim5\times10^{-3}$~days, $\nu_{\rm m,RS}$ is in the $r$-band around this time ($\approx10^{-3}$~days). However, this produces an impossibly low initial bulk Lorentz factor ($\Gamma_0$) for the jet. Taking the most extreme scenario of $\nu_{\rm m,FS}$ and $\nu_{\rm m,RS}$ passing through $r$-band at the latest and earliest possible times, $\approx10^{-2}$~days and $\approx4\times10^{-4}$~days respectively (pushing them apart to the greatest degree), the initial bulk Lorentz factor\footnote{Defined as $\Gamma_0\approx (\nu_{\rm m,FS}(t_{\rm dec})/\nu_{\rm m,RS}(t_{\rm dec}))^{1/2}$ where $t_{\rm dec}\approx 4\times10^{-4}$~days and $\nu_{\rm m,FS}\propto t^{-3/2}$.} $\Gamma_0$ is $\lesssim11$. At the same time, the FS Lorentz factor, $\Gamma_{\rm FS}\approx150$ for $E_{\rm K,iso}\approx 6\times10^{52}$~erg and $n_0\approx0.1$~cm$^{-3}$ (following the closure relations in Section~\ref{sec:GRB180418A_energy_density}). A jet with $\Gamma_0\approx11$ cannot set up an FS with a Lorentz factor of $\Gamma\approx150$. Hence it is unlikely that the relatively shallow optical light curve at $\lesssim 5\times10^{-3}$~days is due to the passage of $\nu_{\rm m,RS}$.

An alternate possibility is to relax the assumption of $g\approx2.2$. Higher values of $g$ have been inferred for long-duration GRBs in the past, with $g\approx5$ for GRB~130427A \citep{Laskar2013,pcc+14} and $g\approx2.7$ for GRB~181201A \citep{lves+19}, each greater than the expected value of $g\approx1$ for a wind-like environment \citep{zwd05}. In our case of GRB\,180418A, we find that $\alpha_{\rm r}\approx-1.8$ for $g\approx5$. While this is still too steep to completely explain the observed decline rate, the addition of the FS component ameliorates the remaining tension. 
For $z=1.0$, we find that an RS+FS model gives consistent parameters that we derived for the FS alone, for $p$, the explosion properties and microphysical parameters. In addition, we find $t_{\rm dec}\approx3.8\times10^{-4}$~days, $\nu_{\rm m,RS}(t_{\rm dec})\approx10^{13}$~Hz, and $\nu_{\rm c,RS}(t_{\rm dec})\approx6\times10^{18}$~Hz fits the multi-frequency data self-consistently. In Figure~\ref{fig:180418a_multimodel} we present our best-fit light curves using the combined RS+FS model with parameters $p=2.4$, $\epsilon_{\rm e}=0.13\zeta$, $\epsilon_{\rm B}=10^{-4}\zeta^{-3}$, $n_0=0.1\zeta^5$~cm$^{-3}$ and $E_{\rm K,iso}=6.2\times10^{52}\zeta^{-2}$~erg. The parameter $\zeta=\frac{1+z}{2}$ analytically encapsulates the additional degeneracy in these parameters due to the unknown redshift. 
For this model, we find a RS magnetization of $R_{\rm B}\equiv\sqrt{\epsilon_{\rm B,RS}/\epsilon_{\rm B,FS}}\approx5.2$ and an initial jet Lorentz factor, $\Gamma_0\approx150\approx\Gamma_{\rm FS}(t_{\rm dec})$, which is commensurate with a non-relativistic RS. Although the bulk Lorentz factor we obtain is similar to that reported by \cite{Becerra2019}, $\Gamma_0\approx160$, our RS magnetization parameter and $E_{\rm K,iso}$ are below and above, respectively, of the reported values by these authors\footnote{We believe that there may be a typographical error in \cite{Becerra2019}, where the reported value of $E_{\rm K,iso}$ is incorrect by a factor of 10. Using their values of the other parameters, we infer $E_{\rm K,iso}\approx6\times10^{51}$~erg would be required to match the X-ray and optical light curves.} ($R_{\rm B}\approx14$ and $E_{\rm K,iso}\approx0.77\times10^{51}$~erg).

\section{X-ray Afterglow Comparison}
\label{sec:GRB180418A_ComparisonPopulation}

%
\begin{figure*}
	\centering
	$\vcenter{\hbox{\includegraphics[width=0.5\textwidth]{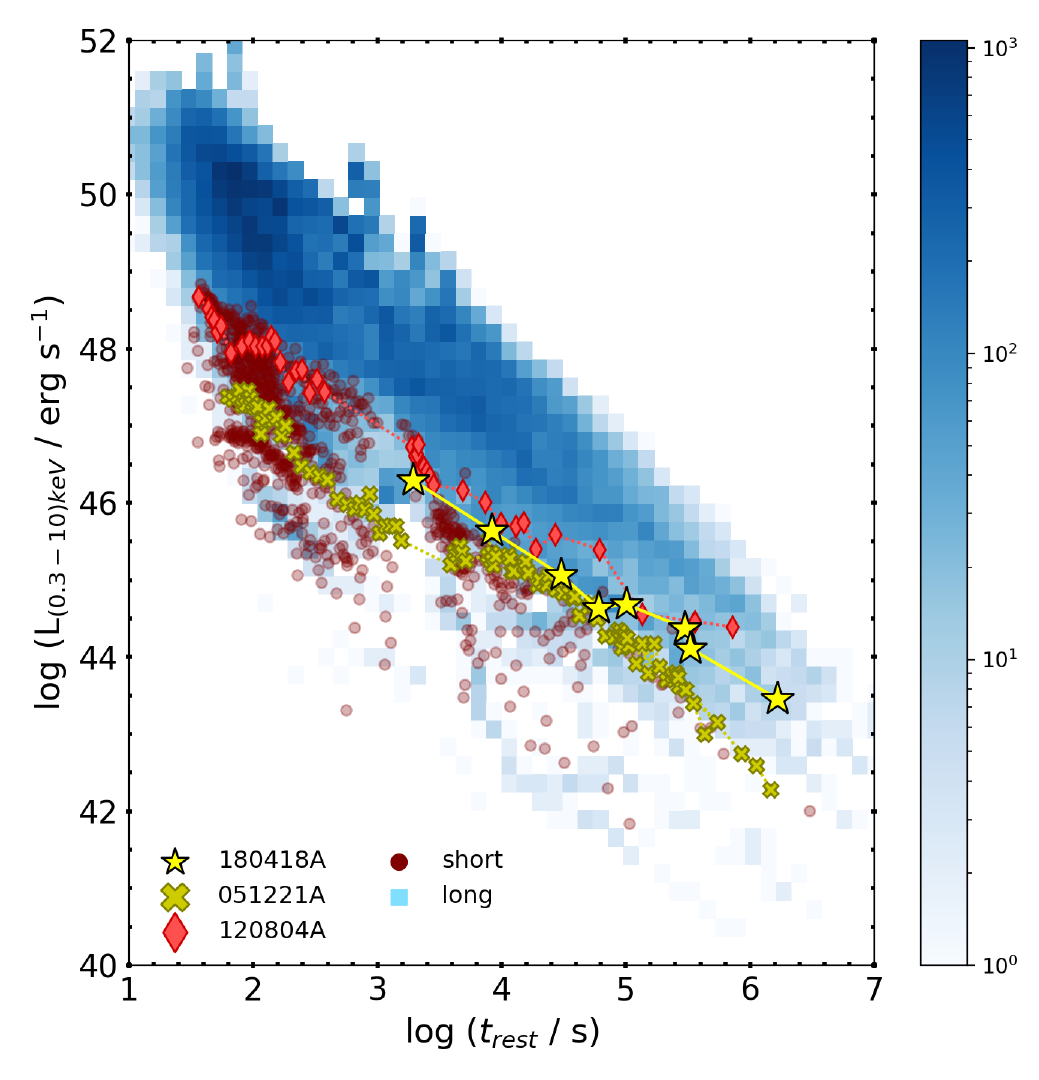}}}$
\hspace*{.1in}
$\vcenter{\hbox{\includegraphics[width=0.45\textwidth]{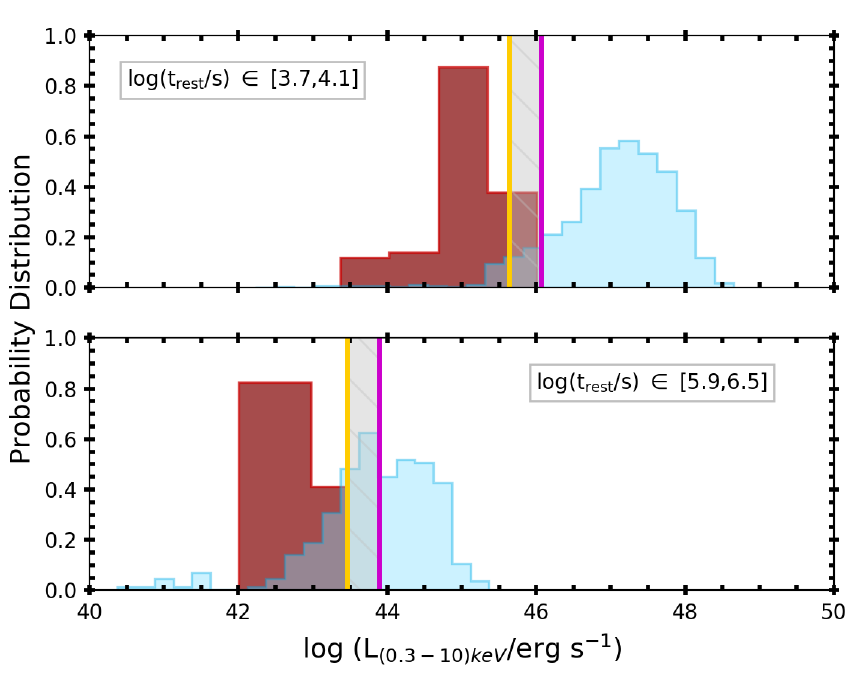}}}$
    \caption{{\it Left:} The X-ray luminosity ($0.3-10.0$\,keV) versus rest-frame time plot of all the GRBs detected by \textit{Swift}/BAT with known redshifts. The long GRB population (T$_{90}>2$\,s) is shown in blue, where the different shades in color represents the density of available data. The short GRB population (T$_{90}\leq2$\,s) is represented by deep red circles. For plotting purposes, we only show the X-ray luminosity light curve of GRB\,180418A at $z=1$ (yellow stars). Those short GRBs that display similar X-ray behavior to GRB\,180418A  (Section~\ref{sec:GRB180418A_ComparisonPopulation}) are shown with different point markers (yellow x's and red diamonds). {\it Right:} Distribution of the X-ray luminosity ($0.3-10.0$\,keV) for the short (red) and long (blue) GRB populations at $\log(\delta t_{\rm rest} /{\rm s}) \approx 4$ (top panel) and $\log(\delta t_{\rm rest} /{\rm s}) \approx 6$ (bottom panel). The grey shaded area indicates the potential X-ray luminosity values for the GRB\,180418A afterglow considering redshift values between $z=1$ and $z=1.5$. The X-ray luminosity of the afterglow at $z=1$ is shown with dark yellow vertical lines, while the dark pink vertical lines indicate the luminosity at $z=1.5$.}
    \label{fig:180418a_LumVsTime}
\end{figure*}
%

In the following section, we compare the X-ray afterglow behavior of GRB\,180418A to the {\it Swift} short and long GRB populations, by performing a systematic comparison of their 0.3-10~keV \textit{Swift}/XRT luminosities ($L_{\rm X}$) and temporal behavior. We obtain the XRT flux light curves \citep{Evans2007,Evans2009} for the GRBs with known redshifts, resulting in 37 short GRBs and 350 long GRBs. We also include the late-time ($\log(\delta t_{\rm rest} / {\rm s}) \gtrsim 5$) \textit{Chandra} and \textit{XMM-Newton} data in the light curves if available, i.e. for GRBs 051221A \citep{Burrows2006}, 120804A \citep{Berger2013}, 150101B \citep{Fong2016}. We calculate the $L_{\rm X}$ and rest-frame times ($t_{\rm rest}$) for each GRB, and plot the light curves in Figure~\ref{fig:180418a_LumVsTime} (\textit{left}). To compare the X-ray afterglow behavior of GRB\,180418A with the short and long GRB populations, we calculate the characteristic median and 1$\sigma$ dispersion values for both populations: $\log(L_{\rm X, short} / {\rm erg\,s^{-1}})=45.18^{+0.51}_{-0.21}$ and $\log(L_{\rm X, long} / {\rm erg\,s^{-1}})=47.14^{+0.84}_{-0.63}$ at $\log(\delta t_{\rm rest} /{\rm s}) \approx 4$, and $\log(L_{\rm X, short} / {\rm erg\,s^{-1}})=42.60^{+0.41}_{-0.42}$ and $\log(L_{\rm X, long} / {\rm erg\,s^{-1}})=43.90^{+0.70}_{-0.69}$ at $\log(\delta t_{\rm rest} /{\rm s}) \approx 6$.

Assuming a redshift of $z=1$ for \grb, we find that the X-ray afterglow luminosity at earlier times ($\log(\delta t_{\rm rest} / {\rm s}) \approx 4$) of $\log(L_{\rm X} / {\rm erg\,s^{-1}}) \approx 45.6$, is sub-luminous compared to the majority of long GRBs, falling $2.5\sigma$ below the median $L_{\rm X}$ of this population at this time, but it is just above the short GRB median, and within the $1\sigma$ uncertainty (Figure~\ref{fig:180418a_LumVsTime}). On the other hand, at late times ($\log(\delta t_{\rm rest} / {\rm s}) \approx 6$), we find that the X-ray luminosity of GRB\,180418A, $\log(L_{\rm X} / {\rm erg\,s^{-1}}) \approx 43.5$, is within the $1\sigma$ uncertainty region of the long GRB population, almost $2\sigma$ above the median of the X-ray luminosity of short GRBs (Figure~\ref{fig:180418a_LumVsTime}, \textit{right}). However, for short GRBs, there exists very sparse information at these late epochs due to their faintness, and in fact the majority of all available information comes from \textit{Chandra} and \textit{XMM-Newton} observations. We find similar result assuming $z=1.5$ (Figure~\ref{fig:180418a_LumVsTime}, \textit{right}).

It is useful to explore the properties of the subsets of long and short GRBs which exhibit similar X-ray light curve behavior to GRB\,180418A. To determine the subsets that track the X-ray afterglow behavior of GRB\,180418A, we select those events with detections within a log-spaced interval of 5\% of the GRB rest-frame $\delta t$ and X-ray luminosity for $\log(\delta t_{\rm rest} / {\rm s}) \leq 5.6$. This interval was chosen to represent the temporal behavior probed by \grb, while also optimizing the number of GRBs in each sample which fit this criteria. Our criteria are satisfied for 2/37 short GRBs and 103/350 long GRBs. If we consider a fiducial value for the redshift of $z=1.5$, our criteria are not satisfied for any short GRB.

The two short GRBs with similar behavior to GRB\,180418A are GRBs\,051221A \citep{Parsons2005a} and 120804A (\citealt{Lien2012}; dark yellow 'x' markers and red diamonds, respectively, in Figure\,\ref{fig:180418a_LumVsTime}). Comparing the $\gamma$-ray properties (duration, hardness ratios and fluence), redshifts, and host properties, we find that these bursts span the full range of short GRBs \citep{Fong2015,Lien2016}. Compared to the other two bursts, GRB\,180418A has the longest duration, with $T_{90}\approx1.9$\,s and is potentially one of the furthest ($z=$\,1-2.25), although we note that the photometric redshift of GRB\,120804A is $z\sim 1.3$ \citep{Berger2013}.

For the subset of 103 long GRBs which are similar in X-ray behavior to GRB\,180418A, the main properties as determined by {\it Swift}/BAT are fairly heterogeneous. However, we find that five of these long GRBs (GRBs\,050416A, 051016B, 090927, 100816A and 140710A; \citealt{Cenko2005,Parsons2005b,Grupe2009,Oates2010,Siegel2014} respectively)\footnote{We note that GRB\,050416A and GRB\,090927 are clear cases of long GRBs since a supernova remnant was detected for GRB\,050416A \citep{Soderberg2007} and GRB\,090927 is most likely a collapsar \citep{Nicuesa2012}.} have $T_{90}<4$\,s, while only 10~GRBs in the entire sample of 350 long GRBs have such durations. This means that half of the available population of long GRBs with $T_{90}\approx 2-4$\,s share X-ray afterglow luminosities and behavior similar to GRB\,180418A. Like GRB\,180418A, this subset falls 2.5$\sigma$ below the long GRB median at early times and within 1$\sigma$ of the median value at late times. To investigate the random chance of detecting long GRBs with T$_{90}<4$\,s, we draw 103 durations from the sample of 350 {\it Swift} long GRBs, 10000 times. We find that in 9\% of cases, we obtain a sample containing 5 GRBs with T$_{90}<4$\,s. If we include GRB\,180418A as part of this sample (making 6/11 of bursts with $T_{90}\leq 4$\,s), this drops to 5\%. Therefore, given the existing duration distribution, we cannot rule out the possibility that the observed statistics are based on random chance. However, the observed trends with X-ray luminosity are nonetheless intriguing, and a correlation between shorter durations and low X-ray luminosity may exist in the long GRB population. We note that these long GRBs are not necessarily the least luminous (Figure~\ref{fig:180418a_LumVsTime}; \citealt{Dereli2017}), but represent those that track the X-ray behavior of GRB\,180418A.

\section{Discussion}\label{sec:GRB180418A_discussion}

\subsection{GRB\,180418A in the $E_{\gamma, \rm peak,i}$--$E_{\gamma,{\rm iso}}$ relation}\label{sec:GRB180418A_short_nature}

\begin{figure}
	\centering
	\includegraphics[width=1.0\columnwidth]{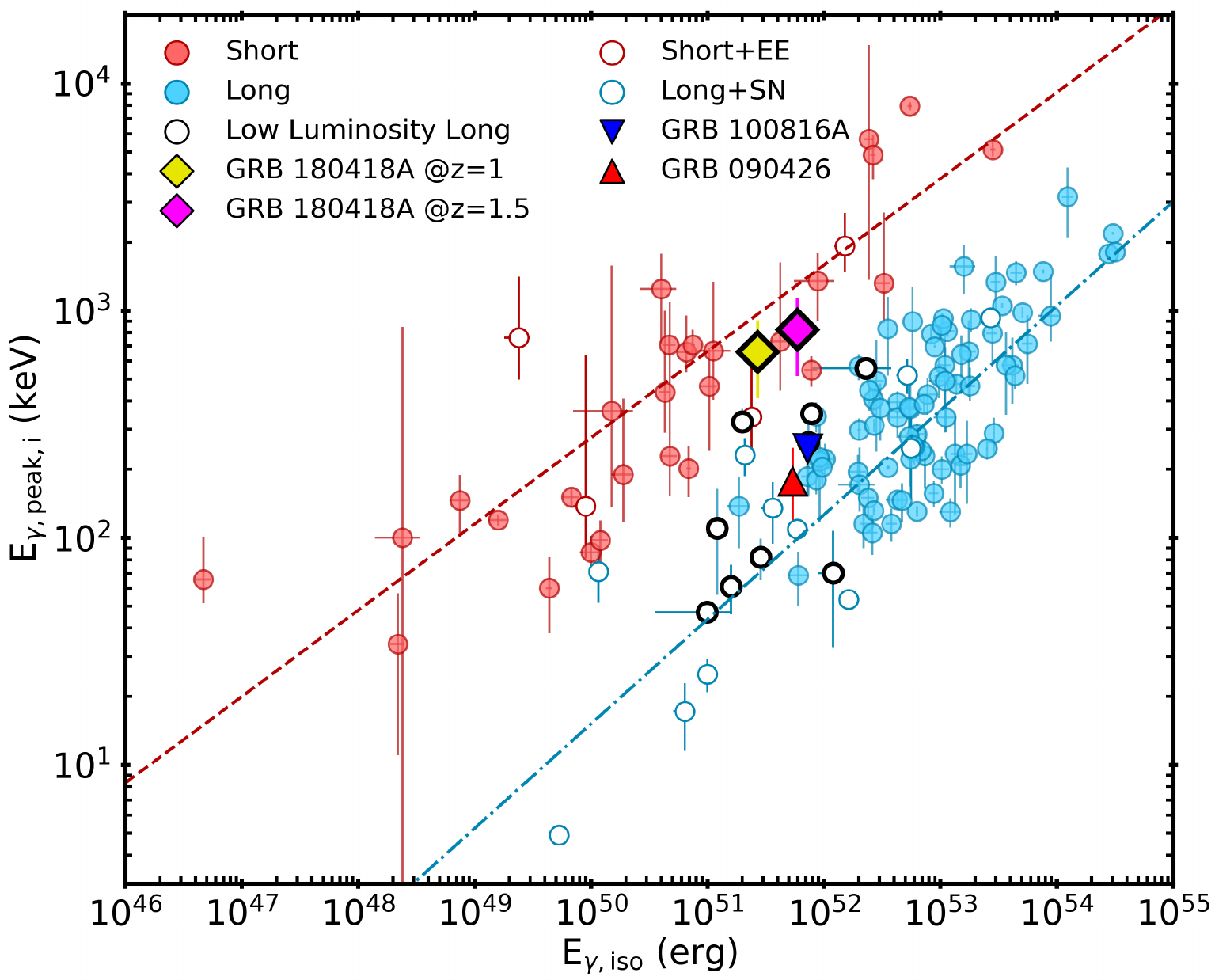}
	\vspace{-0.1in}
    \caption{The $E_{\gamma, \rm peak,i}$-$E_{\gamma,{\rm iso}}$ relation for {\it Swift} and {\it Fermi} short (light red circles) and long (blue circles) GRBs \citep{Minaev2020b}, along with GRB\,180418A (diamonds) at $z=1$ and $z=1.5$. We highlight short GRBs with extended emission (red open circles), and long GRBs with detected supernovae (blue open circles). GRB\,180418A is fully consistent with the short GRB population in the $E_{\gamma, \rm peak,i}$-$E_{\gamma,{\rm iso}}$ plane and is a clear outlier compared to the space occupied by long GRBs (and the Amati correlation). Best-fit correlation models are indicated with lines \citep[red dashed line for short GRBs and blue dashed-dotted line for long GRBs;][]{Minaev2020b}. We also highlight the ambiguous cases of GRBs\,090426 \citep[red triangle;][]{Antonelli2009} and 100816A \citep[blue triangle;][see Section\,\ref{sec:GRB180418A_short_nature}]{DAvanzo2014}, and the low luminosity long GRBs \citep[circles with black borders;][]{Dereli2017}.}
    \label{fig:180418a_Amati}
\end{figure}

From our analysis of GRB\,180418A in the context of the $T_{90}$-hardness plane, we found that the probability of GRB\,180418A being short is 60\% (Section~\ref{sec:GRB180418A_classification}), and that the low density environment is more similar to those inferred for short GRBs. To further elucidate the nature of GRB\,180418A, we compare the spectral properties of its prompt emission to those of short and long GRBs. Several studies \citep[e.g.,][]{Amati2002,Amati2008,Yonetoku2004,Ghirlanda2015} have shown that the energy and luminosity of GRBs follow a correlation; in particular, we explore the correlation (the so-called `Amati relation' \citealt{Amati2002,Amati2008}) between the $E_{\gamma, \rm iso}$ (1--10000\,keV range) and the intrinsic peak energy (i.e., the rest-frame peak energy, $E_{\gamma, \rm peak,i}=E_{\gamma, \rm peak}(1+z)$). Short and long GRBs track different positive correlations in the $E_{\gamma, \rm peak,i}$--$E_{\gamma,{\rm iso}}$ parameter space. Although the nature of this correlation is unclear, it may be connected to the different progenitor channels for both GRB populations, or potentially to viewing angle effects (if the angle between the jet axis and the line of sight of the observer is very small, the harder and brighter the $\gamma$-ray emission will be). The correlation followed by the short GRBs lies above and towards lower $\gamma$-ray energies than the one found for long GRBs, since the $E_{\gamma, \rm peak,i}$ of the short GRBs are generally higher than those of the long bursts (Figure\,\ref{fig:180418a_Amati}).

In the case of GRB\,180418A, we use our \textit{Fermi}/GBM results (Section\,\ref{sec:GRB180418A_classification}) to place the event in the $E_{\gamma, \rm peak, i}-E_{\gamma, \rm iso}$ plane. We find that GRB\,180418A lies closer to the Amati correlation followed by the short GRB population \citep{Minaev2020b}. Indeed, it falls within the space defined by this short GRB class in the $E_{\gamma, \rm peak, i}-E_{\gamma, \rm iso}$ plane, and clearly falls off the correlation for long GRBs (Figure~\ref{fig:180418a_Amati}). In addition, no other long GRBs are consistent with the location of GRB\,180418A. This comparison highlights the similarity in the prompt emission energetics between GRB\,180418A and the short GRB population, pointing towards a possibly-short GRB classification for GRB\,180418A, and supporting our initial expectations. We also compare the $E_{\gamma, \rm peak, i}$ and $E_{\gamma, \rm iso}$ of GRB\,180418A with the \textit{Swift} low-luminosity long GRBs\footnote{We note that $E_{\gamma, \rm peak, i}$ and $E_{\gamma, \rm iso}$ of GRB\,140710A are not available in the literature}, since they do not follow the canonical correlation of long GRBs \citep{Dereli2017}. We also highlight those GRBs of questionable classification (Figure\,\ref{fig:180418a_Amati}): GRB\,090426 \citep{Antonelli2009,Levesque2010} with $T_{90}\approx1.25$\,s and similar prompt emission spectral properties, energy scales, and host properties to long GRBs, and GRB\,100816A \citep{DAvanzo2014} with $T_{90}\approx2.9$\,s, one of the long GRBs with similar X-ray behaviour to GRB\,180418A (see Section\,\ref{sec:GRB180418A_ComparisonPopulation}), that was initially classified as a short GRB by \cite{Norris2010}. We note that these events are not consistent with GRB\,180418A within the errors, and that GRB\,180418A does not appear to be an ambiguous case in terms of its placement on the Amati relation in the short GRB class (Figure\,\ref{fig:180418a_Amati}).

With the borderline $\gamma$-ray duration of GRB\,180418A, it is also worth exploring how it compares to the proposed group of intermediate-duration GRBs. While the small group of intermediate-duration GRBs tends to populate the $E_{\gamma, \rm peak, i}-E_{\gamma, \rm iso}$ correlation of long GRBs \citep{deUgarte2011,Horvath2006}, GRB\,180418A does not clearly fall in this class. However, we note that some of the GRBs classified as intermediate-duration events by \cite{deUgarte2011} have been later identified as short GRBs with extended emission (e.g., GRBs 050724 and 060614) or long GRBs with detected supernovae (e.g., GRBs 050416A and 081007; \citealt{Minaev2020b}).

%
%

\subsection{GRB\,180418A environment and reverse shock}\label{sec:GRB180418A_galactic environment}

The detection of the afterglow of GRB\,180418A not only enables us to investigate its burst properties, but also its local and galactic environment. Our Gemini observations revealed a faint host galaxy for GRB\,180418A at an angular offset of $\delta R = 0.16 \pm 0.04''$. Although a secure redshift for GRB\,180418A is not known, the inferred value of the GRB\,180418A host luminosity, $L\approx$\,0.01-1$L^*$ over the presumed redshift range of $z\approx$\,1-2.25, is more consistent with the sub-$L^*$ host galaxies of long GRBs \citep{Savaglio2009,Blanchard2016} than the hosts of short GRBs which are typically at $0.5-3L^*$ \citep{Berger2014a,Paterson2020}. The angular offset translates to a projected physical distances of $1.29\pm0.33$\,kpc at $z=1$ ($1.38\pm0.34$\,kpc at $z=1.5$). This places the burst at the lower end of the projected physical offsets range for short GRBs, closer to its host than 90\% of the known short GRBs \citep{FongBerger2013}. Considering the long GRB population, which has smaller projected physical offsets, GRB\,180418A falls at the median of the population \citep{Blanchard2016}. Given the proximity of the event to the host center, it is less expected, however, to find the low inferred circumburst density values that we do for GRB\,180418A, $\approx 10^{-2}-10^{-4}$\,cm$^{-3}$, which are more consistent with the inferred values of short GRB circumburst environments. Since we are considering projected physical distances, there is still a possibility (since we are missing the depth component) for the real distance of GRB\,180418A from the center of its host to be larger and, therefore, explaining the low density values inferred for this event.

We note that for six long GRBs with clearly detected RSs \citep[GRBs\,990123, 130427A, 160509A, 161219B, 160625B and 181201A;][]{Meszaros1999,Laskar2013,Laskar2016,Laskar2018a,Alexander2017,lves+19} the circumburst densities are very low, $\approx 5 \times 10^{-5}$--$10^{-2}$\,cm$^{-3}$ \citep{Laskar2018a}. In the case of short GRBs, there are three events for which radio detections of RSs have been claimed: GRBs\,051221A \citep{Soderberg2006}, 160821B \citep{Lamb2019,Troja2019}, and potentially for GRB\,200522A \citep{Fong2020}. For these events, the inferred circumburst densities are low although more consistent with average short GRBs, ranging between $\approx 10^{-4}$-$10^{-2}$\,cm$^{-3}$. In this framework, it is thought that these low density environments, for both long and short GRBs, may be responsible for a slow cooling reverse shock, which allows the RS emission to be detectable for longer \citep{clf04,Laskar2013}.
In the case of GRB\,180418A, the circumburst values we inferred (Table~\ref{tab:GRB180418A_energy_density}) are in agreement with those seen in the RS scenario. 

Including RS emission potentially explains the excess in the early-time ($\delta t \approx 1$\,day) afterglow emission of GRB\,180418A (\citealt{Becerra2019}; and this paper). If GRB\,180418A is indeed a short GRB, then it will be the first with a RS detected in the optical band and with self-consistent RS model parameters ($\Gamma_0\gtrsim150$ and $R_{\rm B}\approx5.2$; Section\,\ref{sec:GRB180418_RSFS_scenario}). The other short GRB with reported values of the initial bulk Lorentz factor and magnetization parameter is GRB\,160821B \citep{Lamb2019}, however \cite{Lamb2019} inferred these values from the FS parameters instead of using the information from the RS spectral parameters as we do in our work. In the case of GRB\,051221A and GRB\,200522A, the jet Lorentz factors are between $18-26$ and $\gtrsim10$ respectively \citep[][]{Soderberg2006,Fong2020}, but in both studies, assumptions on the magnetization parameter were made. On the other hand, comparing the GRB\,180418A reverse shock parameters with those of the long GRBs ($\Gamma_0\approx100-300$ and $R_{\rm B}\approx0.5-10$; \citealt{Laskar2018a}), we find that the values for GRB\,180418A are encompassed by the ranges of the initial jet Lorentz factor and magnetization parameter of long GRBs.

\subsection{GRB\,180418A and jet opening angles}\label{sec:GRB180418A_jet}

%

\begin{figure*}[!t]
\centering
\includegraphics[width=0.49\textwidth]{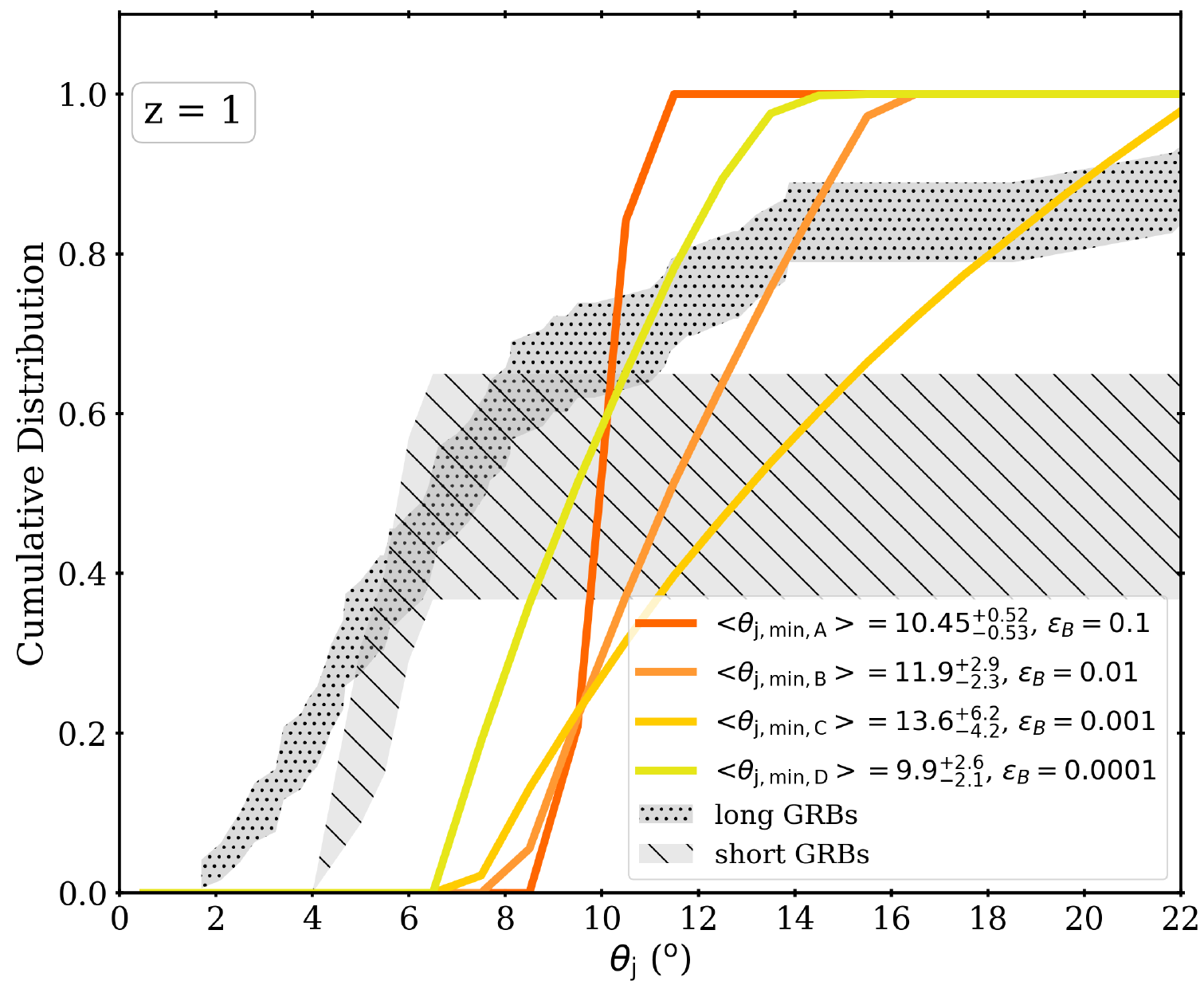}
\includegraphics[width=0.49\textwidth]{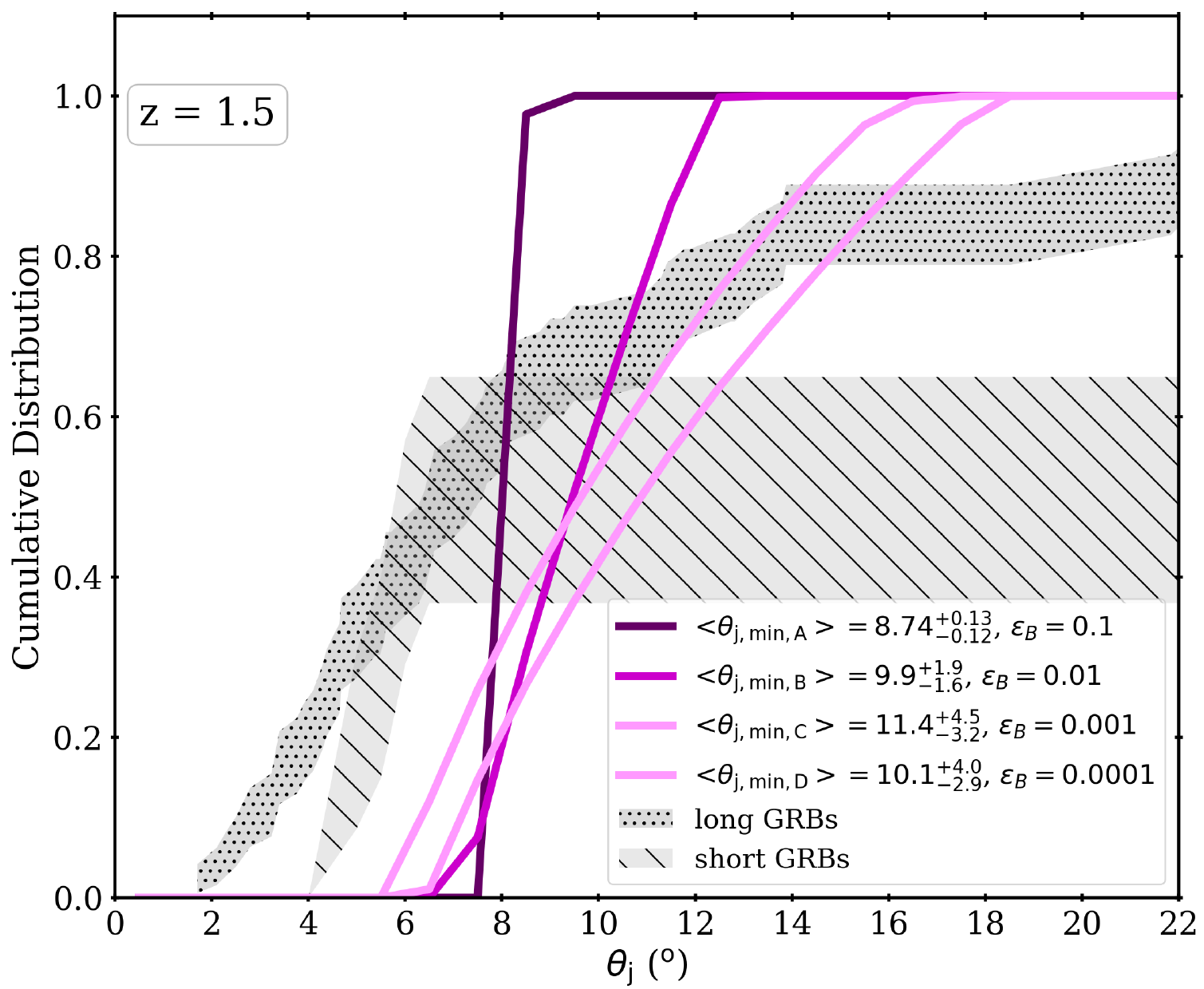}
\vspace{-0.1in}
\caption{The cumulative distributions of the minimum values of the jet opening angles ($\theta_{\rm j}$) at $z=1$ (\textit{left}) and $z=1.5$ (\textit{right}) for GRB~180418A. The different colors correspond to the different cases shown in Table~\ref{tab:GRB180418A_energy_density}. The cumulative distributions of the opening angles for short and long GRBs are shown in the dashed light-grey and dotted dark-grey areas, respectively. In this case, we have applied survival statistics for right-censored data to account for the lower limits of the opening angles in both populations. The plotted areas correspond to their 68\% confidence intervals.
\label{fig:180418a_OpeningAngle}}
\end{figure*}
%

Finally, we compare the cumulative distributions of the minimum values for opening angles of \grb\ with those for the opening angles corresponding to the short and long GRB populations (see Figure~\ref{fig:180418a_OpeningAngle}). From a progenitor standpoint, the massive star progenitors of long GRBs provide a natural collimating medium prior to jet break-out \citep{MeszarosRees2001,Zhang2003}. In contrast, BNS and neutron star-black hole (NS-BH) mergers have no clear analogous mechanism to maintain jet collimation beyond the jet's breakout from the kilonova ejecta. Based on jet predictions of simulations of post-merger black hole accretion \citep[][]{Ruffert1999,Aloy2005,Rezzolla2011}, the general expectation is that short GRBs can achieve wider jet opening angles. The wider jets, coupled with their circumburst density environments, which are orders of magnitude lower than long GRBs \citep{Fong2015}, lead to later expected break times (see Equation~\ref{eqn:jb}). Taken together, these characteristics present an observational challenge in detecting signatures of the expected wider jets in short GRBs. Indeed, our knowledge of short GRB jets generally comes from a few measurements with $\theta_{\rm j} \approx 6^{\circ}$ \citep{Fong2015}.

With the non-detection of a jet break to $\approx 38.5$~days in the X-ray afterglow of GRB\,180418A, we infer an opening angle constraint of $\theta_{\rm j} \gtrsim 9-14^{\circ}$ (Figure~\ref{fig:180418a_OpeningAngle}), depending on the value of the redshift, microphysical parameters and explosion properties. 
For instance, from the best-fit multi-wavelength model, we find $\theta_{\rm j}\gtrsim17.3^{\circ}\zeta^{1/2}$. 
This limit constrains the jet of GRB\,180418A to be relatively wide in the context of the distribution of long GRBs, which have a median opening angle of $\theta_{\rm j} \approx 7^{\circ}$ (Figure~\ref{fig:180418a_OpeningAngle}; \citealt{Frail2001,Bloom2003,Goldstein2016}), and $\approx 75\%$ of which have $\theta_{\rm j} \lesssim 10^{\circ}$. Instead, the opening angle constraint of GRB\,180418A is more consistent with the short GRB distribution, which only consists of six jet measurements and several lower limits to date with $\langle \theta_{\rm j} \rangle = (16\pm10)^{\circ}$ \citep{Fong2015}. GRB\,180418A increases the small sample of GRBs with wide opening angle constraints, in particular, GRB050724A with $\theta_{j}\gtrsim 25^{\circ}$ \citep{Berger2005c},  GRB\,120804A with $\theta_{j} \gtrsim 13^{\circ}$ \citep{Berger2013,Fong2015} and GRB\,150101B with $\theta_{\rm j} \gtrsim 9^{\circ}$ \citep{Fong2016}, all inferred from late-time X-ray observations. 

We can also use the opening angle to calculate the {\it lower} limit on the beaming-corrected, true energy scale of GRB\,180418A to be $E_{\rm true, tot} \equiv E_{\rm K}+E_{\gamma}\gtrsim(1.97-44)\times 10^{50}$\,erg at $z=1$, and $E_{\rm true, tot}\gtrsim(2.68-57)\times 10^{50}$\,erg at $z=1.5$, with the corresponding upper limits set by the isotropic-equivalent total energies of $E_{iso, tot}=(1.2-29)\times10^{52}$\,erg and $E_{iso, tot}=(2.2-36)\times10^{52}$\,erg. The mechanisms that power the relativistic jet \citep{Shibata2019}, either the thermal energy that is released during the $\nu\bar{\nu}$ annihilation process in baryonic outflows \citep{Jaroszynski1993,Mochkovitch1993}, or magnetohydrodynamic (MHD) processes in the accretion remnant of a black hole \citep[e.g.,][]{Blandford1977,Rosswog2003,Ruiz2016,Siegel2017}, are expected to attain different energy releases. In particular, it is expected that the released energy from $\nu\bar{\nu}$ annihilation mechanism reaches levels of $10^{48-49}$\,erg \citep{Birkl2007}, with larger energy scales of $>10^{50}$\,erg for magnetized jets \citep[e.g.,][]{Blandford1977,Rosswog2003,Ruiz2016,Siegel2017}. In addition, theoretical studies have shown that there are different jet opening angle predictions based on the magnetization of the jet \citep{Rosswog2002,Duffell2018,Nathanail2020}, as well as different outcomes for BNS and NS-BH mergers \citep{Murguia-Berthier2017}, with more magnetized outflows found to produce wider jets with $\theta_{\rm j}\gtrsim 10^{\circ}$ \citep{Nathanail2020}.

\section{Conclusions}\label{sec:GRB180418A_conclusions}

In this paper we present the multi-wavelength monitoring campaign on the afterglow of the possibly-short GRB\,180418A and the discovery of its faint host galaxy. In particular, the superb angular resolution of {\it Chandra} allowed us to disentangle a contaminating source in the {\it Swift}/XRT aperture, and track the afterglow to $\delta t \approx 38.5$~days. Our main conclusions are summarized as follows:

\begin{itemize}
    \item In terms of traditional classification schemes such as the $T_{90}$-Hardness plane and the Amati relations, we find that GRB\,180418A is more likely a short GRB. In the context of the {\it Fermi}/GBM population, we find a probability of being short (from the $T_{90}$-Hardness plane) of $60\%$, and consistency within the population of short GRBs in the Amati relation.
    \item The detection of the X-ray afterglow at $\delta t \approx 38.5$ days makes this burst one of the very few short GRBs with a late-time detection in X-rays ($\gtrsim$ 20~days).
    \item The X-ray afterglow light curve, coupled with the optical multi-band detections, exhibits a single power-law decline. We calculate the lower limit of its jet opening angle to be $\theta_{\rm j}\gtrsim9-14^{\circ}$ (assuming $z=1-1.5$). These lower limits reveal a moderately wide jet angle that is consistent with the distribution of angles for short GRB jets and the expectations for BNS/NS-BH merger relativistic outflows.
    \item When comparing the X-ray afterglow luminosity of GRB\,180418A with those of the short and long GRBs detected by \textit{Swift}/BAT, we find that only two short GRBs track the behavior of GRB\,180418A. We also notice that half of the available population of long GRBs with T$_{90}\approx 2-4$\,s show X-ray afterglow luminosities and behavior similar to GRB\,180418A.
    \item Modeling the afterglow with a joint synchrotron forward and reverse shock, we find beaming-corrected energy scales of $E_{\rm true,tot}\gtrsim(1.97-44)\times10^{50}$\,erg and $E_{\rm true,tot}\gtrsim(2.68-57)\times10^{50}$\,erg, and circumburst densities of $n_{0}=(2.56-110)\times10^{-4}$\,$cm^{-3}$ and $n_{0}=(2.21-160)\times10^{-4}$\,$cm^{-3}$ at $z=1$ and $z=1.5$, respectively. The low inferred circumburst density is also consistent with both short and long GRBs with detected reverse shocks.
    \item GRB\,180418A is the first short GRB with a reverse shock detected in the optical band with self-consistent RS model parameters.
    \item We find a faint host galaxy coincident with the {\it Chandra} X-ray and optical afterglow positions. The featureless afterglow and host spectrum, coupled with the detection of the afterglow with UVOT, constrain the redshift range of the burst to most likely be $z \approx$\, 1-2.25.
\end{itemize}

The continuous coverage in the optical and the late-time detections in X-rays, coupled with the nature of GRB\,180418A, make this event an exceptional GRB case. Our work demonstrates that multi-wavelength afterglow observations are essential not only at early times following the GRB trigger (detection of reverse shock in the optical; \citealt{Becerra2019}), but also at late times (better constraints of the jet opening angle). The power of ToO multi-wavelength campaigns is vital for further investigating the increasingly diverse behavior of GRB afterglows, determining the energetics and environments where bursts occur, and studying the potential GRB central engine and progenitor channels. More deep follow-up observations are necessary and encouraged to increase the number of detected jet breaks in future short GRB afterglows.

\clearpage

\acknowledgments
The authors acknowledge Phil Evans for the useful advice to obtain the full set of X-ray GRB afterglow light curves from the UK Swift Science Data Centre. The authors thank Daniel Perley for his valuable contribution to proposals.

\noindent The Fong Group at Northwestern acknowledges support by the National Science Foundation under grant Nos. AST-1814782 and AST-1909358. Support for this work was provided by the National Aeronautics and Space Administration through Chandra Award Number G08-19025X issued by the Chandra X-ray Center, which is operated by the Smithsonian Astrophysical Observatory for and on behalf of the National Aeronautics Space Administration under contract NAS8-03060. P.V. acknowledges support from NASA grants 80NSSC19K0595 and NNM11AA01A.

\noindent This work made use of data supplied by the UK Swift Science Data Centre at the University of Leicester.

\noindent Based on observations obtained at the international Gemini Observatory (Program IDs GS-2018A-Q-127, GN-2018A-Q-121, GN-2018B-Q-117), a program of NOIRLab, which is managed by the Association of Universities for Research in Astronomy (AURA) under a cooperative agreement with the National Science Foundation on behalf of the Gemini Observatory partnership: the National Science Foundation (United States), National Research Council (Canada), Agencia Nacional de Investigaci\'{o}n y Desarrollo (Chile), Ministerio de Ciencia, Tecnolog\'{i}a e Innovaci\'{o}n (Argentina), Minist\'{e}rio da Ci\^{e}ncia, Tecnologia, Inova\c{c}\~{o}es e Comunica\c{c}\~{o}es (Brazil), and Korea Astronomy and Space Science Institute (Republic of Korea).

\noindent This work was enabled by observations made from the Gemini North telescope, located within the Maunakea Science Reserve and adjacent to the summit of Maunakea. We are grateful for the privilege of observing the Universe from a place that is unique in both its astronomical quality and its cultural significance.

\noindent Observations reported in this paper were obtained at the MMT Observatory, a joint facility of the University of Arizona and the Smithsonian Institution (Programs 2018C-UAO-G4, UAO-G213-20A, UAO-G4, UAO-G7, UAO-G212-20A, 2018b-UAO-G15). MMT Observatory access was supported by Northwestern University and the Center for Interdisciplinary Exploration and Research in Astrophysics (CIERA).

\noindent Observations obtained by the United Kingdom Infrared Telescope (UKIRT; program: U/18A/UA01) was supported by NASA and operated under an agreement among the University of Hawaii, the University of Arizona, and Lockheed Martin Advanced Technology Center; operations are enabled through the cooperation of the East Asian Observatory. We thank the Cambridge Astronomical Survey Unit (CASU) for processing the WFCAM data and the WFCAM Science Archive (WSA) for making the data available.

\bibliographystyle{aasjournal}
\bibliography{journals_apj,reference}

\vspace{5mm}
\facilities{\textit{FERMI}/GBM, \textit{Swift}(BAT and XRT), \textit{Chandra}(ACIS-S), Gemini-S and -N(GMOS), UKIRT(WFCAM), MMT(MMIRS and Binospec)}

\software{\texttt{IRAF} \citep{Tody1986,Tody1993}, \texttt{SExtractor} \citep{Bertin1996}, \texttt{CIAO} software package \citep[v.4.12][]{Fruscione2006}, \texttt{HEASoft} software \citep[v.6.17;][]{Blackburn1999,NASA2014}, \texttt{HOTPANTS} \citep{Becker2015}, \texttt{mclust} \citep{Scrucca2016}, \texttt{XSPEC} \citep{Arnaud1996}, \texttt{lifelines} \citep{lifelines}}

\end{document}